# Aharonov-Bohm interference and the evolution of phase jumps in fractional quantum Hall Fabry-Perot interferometers based on bi-layer graphene

Jehyun Kim[1], Himanshu Dev[1], Ravi Kumar[1], Alexey Ilin[1], André Haug[1], Vishal Bhardwaj[1], Changki Hong[1], Kenji Watanabe[2], Takashi Taniguchi[3], Ady Stern[1], and Yuval Ronen[1]*

[1] *Department of Condensed Matter Physics, Weizmann Institute of Science, Rehovot, Israel*

[2] *Research Center for Functional Materials, National Institute for Materials Science, Tsukuba, Japan*

[3] *International Center for Materials Nano architectonics, National Institute for Materials Science, Tsukuba, Japan*

\* *Corresponding author*: yuval.ronen@weizmann.ac.il

**Abstract**

Quasi-particles in fractional quantum Hall states are collective excitations that carry fractional charge and anyonic statistics. While the fractional charge affects semi-classical characteristics such as shot noise and charging energies, the anyonic statistics is most notable in quantum interference phenomena. In this study, we utilize a versatile bilayer graphene-based Fabry-Pérot Interferometer (FPI) that facilitates the study of a broad spectrum of operating regimes, from Coulomb-dominated to Aharonov-Bohm, for both integer and fractional quantum Hall states. Focusing on the $\nu = 1/3$ fractional quantum Hall state, we study the Aharonov-Bohm interference of quasi-particles when the magnetic flux through an interference loop and the charge density within the loop are independently varied. When their combined variation is such that the Landau filling remains 1/3 we observe pristine Aharonov-Bohm oscillations with a period of three flux quanta, as is expected from the interference of quasi-particles of one-third of the electron charge. When the combined variation is such that it leads to quasi-particles addition or removal from the loop, phase jumps emerge, and alter the phase evolution. Notably, across all cases, the average phase consistently increases by $2\pi$ with each addition of one electron to the loop.

**Introduction**

As two indistinguishable particles interchange positions in our three-dimensional (3D) universe, a fundamental property of their nature is revealed: their *exchange statistics*. This property divides all known elementary particles into bosons or fermions, where their many-body wavefunction acquires, under an exchange operation, zero or $\pi$ phase shift, respectively. This dichotomy can be broken in two-dimensional (2D) space, where for special phases of matter, novel quasi-particles (excitations) can be realized [1–3]. One such phase is the Fractional Quantum Hall Effect (FQHE), where the many-body wave function of excitations (anyons), unlike fermions and bosons, may exhibit a non-trivial exchange phase (Abelian anyons) [4], or even rotate to a new wavefunction in a highly degenerate ground state subspace (non-Abelian anyons) [5].

A natural way to probe the exchange phase is by interferometry. In a quantum Hall interferometer, the bulk is insulating, yet may host localized anyons, while chiral edge modes propagate along the edges of the sample [6,7]. Bulk-edge correspondence allows the determination of the state's properties (in the bulk) via studying the edge modes. Explicitly, under certain conditions, the Aharonov-Bohm (AB) interference of edge modes can reveal the interfering quasi-particles' charge (via flux periodicity), and even demonstrate the anyonic exchange statistics accumulated by the interfering waves propagating around the localized anyons [8,9].

Electron-based interferometry in the framework of the FQHE has advanced considerably during the last two decades in the GaAs/AlGaAs heterostructure platform grown by Molecular beam epitaxial (MBE) [10–14]. The two main interferometers are the Mach-Zehnder interferometer (MZI) [13,15–17] and the Fabry-Pérot interferometer (FPI) [8,10–12,18–20]. The FPI supports more than two interfering paths; the area of its interference loop may change with the magnetic field in a way that is not fully controlled; and it is affected by Coulomb interactions. The MZI supports two interfering paths; its interfering area is insensitive to the magnetic field and is relatively unaffected by Coulomb interactions; yet its internal drain forces a periodicity of one flux quantum even when the interfering particle carries a fractional charge. Recent interference experiments, performed with FPI and MZI in a GaAs-based system, exhibited promising results [10,12,13,21]. Here, we present results of interfering integer [22–25] and fractional charges in an FPI based on van der Waals-based heterostructures (vdW).

Graphene-based FPIs in the QHE regime enjoy several advantages over the GaAs-based ones. The energy gaps are larger [26–28]; a single heterostructure can be tuned to both odd and even denominator states, and very close screening gates minimize Coulomb interactions. Here, we will introduce a tunable bilayer graphene-based FPI operating in the integer and fractional regimes. First, we demonstrate a tunable interference between the Aharonov-Bohm (AB) and Coulomb-Dominated (CD) regimes in the integer regime. Based on that, we demonstrate an AB interference at filling $\nu = 1/3$, pointing at a quasi-particle charge of $e/3$. Controlling both the magnetic field and the electron density within the interferometer, we may tune between carrying out the experiment along a line of constant filling to a line of constant density. Doing that, we observe the emergence of phase jumps, which may result from added (subtracted) anyonic QPs in the bulk, thus pointing at the unique exchange phase of $e/3$ anyons. Furthermore, we demonstrate a fundamental relation between the interference phase and the number of electrons within the interference loop.

**Design and operations of a bilayer graphene-based FPI**

The FPI was fabricated on a high-mobility bi-layer graphene heterostructure with a mobility $\mu \geq 2.3 \times 10^5 \text{ cm}^2/\text{V}\cdot\text{s}$ (see SI1). Fig. 1a shows a false-color scanning electron microscopy (SEM) image describing the main structure of the FPI. The high electron mobility is achieved by encapsulating the bilayer graphene with hexagonal boron nitride (hBN) insulating layers and top/bottom conductive graphite layers, as shown in Fig. 1b, thereby screening Coulomb interactions, and smoothing external disorder [28,29]. To define the FPI, we divided the top graphite into eight sections (separated by 40nm etched trenches). We tuned the local filling factor ν beneath each part, thereby guiding the edge modes electrostatically. Each section of the top graphite plays an essential role in the operation of the FPI [22]. The fabricated area of the interference loop is $1\mu m^2$.

Fig. 1a inset shows the left, center, and right gates (LG, CG, and RG) that determine the bulk 2DEG filling factor, ν. We define the quantum point contacts (QPCs) electrostatically by guiding the edge modes to close proximity, thereby inducing tunneling between them. Split gates at the left and right QPCs (LSG and RSG) are used to realize the two QPCs forming the FPI. A Plunger gate (PG) is used to carefully alter the trajectory of the QH edge mode, thereby manipulating the interference area. Lastly, two air bridges (LBG and RBG) are used as gates to fine-tune the transmission of each QPC.

Fig. 1b illustrates the FPI heterostructure and measurement setup. The device is placed in a highly filtered dilution-refrigerator at a base temperature of 30 mK and a perpendicular magnetic field, *B*, placing the 2DEG in the QH regime, *e.g.*, operating in ν = 1 or 1/3 filling. Bias current, $I_{SD}$, is injected along the QH edge modes with an anti-clockwise chirality for the electron carrier type. It is important to note that a single ground is employed. To form the FPI, we tune the filling beneath LSG and RSG to the ν = 0 incompressible states, and RG, LG, and CG to ν, thereby forming the left and right QPCs, LQPC and RQPC.

The transmission through the FPI is determined via the diagonal resistance $R_\text{D} = \frac{V_\text{D}^+ - V_\text{D}^-}{I_\text{SD}}$ (see Fig. 1b and methods section). In the case of a single QPC, the diagonal resistance will follow $R_\text{D} = \frac{1}{t_\text{QPC}} \frac{h}{e^2}$ where $h$ is the Planck constant, $e$ is the electron charge, and $t$ is the QPC transmission. Generally, one would expect an $R_D$ of a relatively open FPI to follow $R_D \sim \cos\theta$, for a 'two-path' interference (*i.e.*, the case of two QPCs that are highly transmissive), where $\theta$ is the sum of the AB phase and the exchange phase [8,30,31].

$$\theta = 2\pi \frac{e^*}{e} \frac{\Phi}{\Phi_0} + \Delta N_{qp} \theta_{anyon} \; ; \; \Phi \equiv A \cdot B \; , \tag{1}$$

where $A$ is the interfering area, $\Phi_0 = h/e$ is the flux quantum, and $e^*$ is the charge of the interfering quasi-particle. The first term in this expression is the Aharonov-Bohm phase. Importantly, the variation of the flux $\Phi$ with the magnetic field can be made more complicated by the possible dependence of the area, $A$, on the magnetic field. This dependence distinguishes between the AB and CD regimes. It may be analyzed by means of a 2D Fourier Transform of the interference data[4,8,19,30,31]. The second term is the added statistical phase due to $\Delta N_\text{qp}$ quasi-particles. We expand on the significance of the two terms below, in the context of fractional states, and highlight an alternative viewpoint on equation (1).

**Interference at the IQHE**

We first operate the FPI in the IQHE, where $\frac{e^*}{e} = 1$ and $\theta_{anyon} = 0$. We work with ν = 1, 2 (beneath LG, CG, and RG regions), while keeping ν = 0 beneath the split gates (LSG and RSG regions) thereby forming the two QPCs. In this operating regime, we source a current, $I_\text{SD}$ = 500 pA, which impinges on each of the QPCs (while the other is fully open) with $R_D$ measuring the QPC's individual transmissions, $t_L$ or $t_R$. The QPCs' transmissions are set to ~0.5 via the applied voltage to LBG and RBG, thus tuning the saddle point in the QPCs' potential (see SI2). Under these conditions, $R_D$ is measured as a function of the applied voltage to the plunger gate, $V_{PG}$, and the magnetic field, $B$.

Fig. 2a shows the measured 2D ('pajama' in the $B$-$V_{PG}$ plane) of $\Delta R_D = R_D - \langle R_D \rangle$, where $\langle R_D \rangle$ is the average value of $R_D$, reflecting the interference pattern of the first Landau level (LL1), at ν = 1. Its corresponding 2D-FFT is shown in Fig. 2c, plotted as a function of $\Phi_0/\Delta B$ and $1/\Delta V_{PG}$, clearly indicating that the oscillations are independent of $B$, suggesting strong area variation ("breathing") with the variation of the magnetic field due to Coulomb domination (CD) [8,19,32]. However, as the density in the FPI was increased (via the $V_{CG}$) to ν = 2, while still interfering LL1 (Fig. 2d inset), the interference pattern switched to a clear AB-type oscillation pattern, with a 'negative' slope of the constant phase lines [22–24,33] and a flux periodicity of one flux quantum (Fig. 2b and 2D-FFT at 2d). QPC transmissions are kept at ~0.5 transmission.

The CD and AB regimes in an FPI have been studied widely in the IQHE in GaAs-based FPIs [14,19,20,33,34], and thus can be understood by comparing two main capacitances in the device, defining a figure of merit for the interference regime, $\xi = \frac{C_{eb}}{C_b + C_{eb}}$, where $C_{eb}$ is edge-to-bulk capacitance (bulk being compressible) and $C_b$ is bulk-to-gates capacitance [8]. The pure CD (AB) regime can be defined when $\xi$ approaches 1 (0). The device's capacitances are determined by geometrical factors such as the thickness of hBN, the interfering area defined, etc., as well as quantum capacitances. Raising the $V_{CG}$ potential leads to an increase in the density within the interference loop, eventually crossing from CD to AB regime, either by reducing $C_{eb}$ or increasing $C_b$. This crossover can be attributed due to the change in the screening owing to the additional inner edge state realized when the bulk filling factor is at ν = 2. A similar behavior of the interfering inner mode of the 2nd LL, (LL2 at ν = 2, 3) is shown in Fig. 2e and 2f (see SI3). The slope of the constant phase lines varies from a positive to a negative slope with increasing $V_{CG}$, suggesting a crossover from CD (Fig. 2g) to AB (Fig. 2h) regimes. While in Figs. 2b, 2d a clean AB interference is seen, in Fig. 2f we observe in the 2D-FFT a strong AB peak and a reminiscent CD peak, conveying a small CD contribution which can be seen by the faint phase jumps in Fig. 2f. Fig. 2i shows the measured $\frac{\Phi_0}{\Delta B}$ as a function of $V_{CG}$ (negative values point at CD regime). For both LL1 and LL2, $\frac{\Phi_0}{\Delta B}$ tends to increase monotonically and converge to interfering areas of 1 μm² and 0.8 μm², respectively. This dependence on the voltage applied to the central gate demonstrates the high-tunability of our FPI, operated in both CD and AB regimes. Note that each of the 2D scans requires several hours to complete, showcasing the stability of the heterostructure gates.

**Interference at the FQHE**

The $R_{xx}$ fan diagram, in the $B$-$V_{RG}$ plane (Fig. 3a) shows the well-developed ν=1/3 and ν=2/5 states. We tuned the FPI filling to ν = 1/3 beneath LG, CG, and RG regions while keeping ν = 0 beneath the QPCs split gates. The transmission of the FPI was monitored by measuring $R_D$ with $I_{SD}$=50 pA, with $t_L$ ~ 0.75 and $t_R$ ~ 0.95 with the 2D interference shown in Fig. 3b (see SI2 and SI5). A clean AB-dominated interference pattern is observed and a complementary 2D-FFT is calculated in Fig. 3d. The magnetic field period for the IQHE interference is $\Delta B = 4.1\ mT$ for LL1. In contrast, the period in Fig. 3b for the ν=1/3 FQHE interference is $\Delta B = 11.9\ mT$. The ratio of the measured periodicities, ~0.35, points to the difference between an electron and a quasi-particle charge with a value of $e^* = e/3$ [35,36]; assuming the interference area is kept constant.

An intriguing scenario arises as we increase $V_{CG}$, while keeping ν = 1/3 beneath the reservoirs regions. As the filling in the interfering area reaches ν = 1 the FPI experiences a major change, where the fractional state within the interference loop is replaced by an integer state, see insets in Figs. 3d and 3e, respectively [37,38]. Surprisingly, the AB oscillations (Fig. 3b) transit to CD oscillations (Fig. 3c) similar to that shown in Fig. 2a. The crossover from AB to CD regime can be understood by considering the charging energy $E_c = (e^*)^2/2C_b$ [8] associated with the different interfering charges. As the interfering charge varies from $e$ to $e/3$, $E_c$ decreases by a factor of 9, consequently giving rise to a reduced $\xi$ where AB oscillations are dominated, strengthening the claim that interference of fractional charges is occurring in Fig 3b. Fig. 3f summarizes the 2D oscillations plotted with $\frac{\Phi_0}{\Delta B}$ and $R_{xx}$ as a function of $V_{CG}$ (see SI4), demonstrating that $\xi$ can be tuned not only by modulating the electrostatic environment but also by the change in the interfering charges. As $V_{CG}$ is increased further the interference pattern shifts to an AB-dominated integer interference regime as can be seen in SI4, due to an additional decrease in the charging energy, in similarity to the previous section.

**Tunability between constant filling and constant density**

The analysis of Fabry-Pèrot interference in the fractional regime is made complicated by the inter-connections between the three parameters that determine the phase in Eq. (1): the area of the interference loop, controlled by $V_{PG}$, the magnetic field, $B$, and the number of localized quasi-particles within the interference loop, affected by the voltage on the central gate $V_{CG}$. The combination of the Coulomb interaction and the incompressibility of the FQH states makes both the area and the number of quasi-particles dependent on the magnetic field, in a non-universal way. To address this complication, we measure the interference as a function the three parameters: $B, V_{CG}, V_{PG}$. We present the resistance as a function of $V_{PG}$ and $B$, taken along lines of fixed slope $\alpha \equiv \frac{dB}{dV_{CG}}$. Two notable values of $\alpha$ correspond to measurement at fixed filling factor $\nu = 1/3$ ($\alpha = 45$ T/V) and measurement at fixed density $\alpha \to \infty$. These two trajectories in the $B$-$V_{CG}$ plane are denoted as ①,③ in Fig. 4a ($R_{XX}$ of the right side of the FPI), ② denoting an intermediate value of $\alpha$ ($\alpha = 120$ T/V). Interference measurements along these three slopes are shown in Figs. 4b, 4c, 4d.

When $\nu$ is held fixed, Fig. 4b, the variation of the magnetic field should not add/remove quasi-particles to the interference loop, and the phase should vary linearly with the magnetic flux. Indeed, at a constant filling where the Fermi level is kept at the center of the fractional gap, line ①, shown in Fig. 4b (and 3d), we observe constant linear phase lines with $\Delta B \sim \frac{3\Phi_0}{A}$, highlighting the dominant contribution of the AB phase term - with an absence of phase slips. An intriguing evolution in the interfering pattern is observed as the trajectory shown in Fig. 4a varies from ① to ③, where phase slips emerge as well as an altered slope in the interference pattern, as shown in Fig. 4c and 4d. The voltage $V_{PG}$ values at which the phase jumps occur do not depend on the magnetic field, unlike what is observed for the GaAs FPIs [10,11]. Focusing on the constant density trajectory, ③, (Fig. 4d) where the variation of $B$ is expected to change the number of quasi-particles, $\Delta N_{qp}$, constant phase lines become less dependent on $B$, resembling a CD regime, excluding a narrow region of $B$, 10.47-10.48 T.

The narrow AB region near 10.47 T can be interpreted by a model where $\Delta B_C = \frac{\Delta_{1/3}\Phi_0 C_b}{\nu e^2 e^*}$ describes the region of $B$ in which the quasi-particles are not allowed to access to the bulk due to the competition between the charging energy and the energy gap for quasi-particle excitations [39]. In our FPI, $\Delta B_C \sim 20$ mT at ~10.5 T, a smaller value than that reported in GaAs-based FPI. This may imply a smaller $\Delta_{1/3}$ in our FPI compared to that in GaAs-based FPI or an increased disorder landscape [10,39].

To further investigate the altered interference pattern, we expand the measurements done in Fig. 4b-d, with additional linear trajectories with a wider range of slopes, $\alpha$, all passing through the intersection point (-0.487 V, 10.5 T). For each value of $\alpha$ and a magnetic field $B$ we carry out a one-dimensional Fourier transform of the conductance as a function of $V_{PG}$. Then, for each trajectory, $\alpha$, we plot the phase evolution of the peak amplitude as a function of magnetic field , $\theta(B)$. The resulting phase increases nearly linearly with $B$, see Fig. S13 in SI6. By fitting the resulting plot to a linear function, we extract $\partial\theta/\partial B$ as a function of $\frac{1}{\alpha} = \frac{\partial V_{CG}}{\partial B}$, and the rms fluctuations of $\theta(B)$ around the linear fit, $\chi_{fit}$ (see SI6). We plot both in Fig. 4e.

This procedure results in two important outcomes. First, we find a linear dependence between $\frac{\partial\theta}{\partial B}$ and $\frac{\partial V_{CG}}{\partial B}$. Since the voltage on the central gate, $V_{CG}$, is proportional to the number of electrons $N$ within the interference loop, this proportionality implies that $\theta \propto 2\pi N$. Indeed, Eq. (1) may be written as $\theta = 2\pi N$, where $N$ is the number of electrons within the interference loop, which is the sum of the number of electrons in the pristine $\nu = 1/3$ fluid, $\frac{\phi}{3\phi_0}$, and one third of the number of quasi-particles. By estimating $\frac{\partial N}{\partial V_{CG}} = \frac{C}{e}$, with $C$ being the capacitance calculated from the measured Streda lines and interference loop area, we are able to obtain, $\theta \approx 0.92 \cdot (2\pi N)$, see SI6. Remarkably, we see this measurement as demonstrating that the phase accumulated by a quasi-particle traversing a closed trajectory in a $\nu = 1/3$ droplet is $2\pi$ times the number of electrons in the droplet [40]. Second, we observe that the deviations from linear dependence are minimal at the value of $\alpha$ that corresponds to a fixed filling factor, a value for which the variation of the magnetic field does not lead to the addition or removal of quasi-particles to the interference loop. Finally, to characterize the size of the phase jumps values, expected to result in $2\pi/3$, we analyzed in SI8 and SI9 a few trajectories of non-constant filling. We observed values with the expected 2π/3 value, co-existing with altered values. This was also seen in a previous experimental work and explained via theoretical work considering an area change [8,11].

**Conclusions**

Our study represents a significant step forward in the field of quantum interference of anyons by constructing and measuring a vdW-based FPI within the fractional QHE. Our device harnesses the unique properties of a high-mobility bilayer graphene conductive layer, allowing us to dynamically tune the interference regime within a single LL from Coulomb-Dominated to Aharonov-Bohm interference through precise electrostatic gating. We further explored the intriguing realm of AB interference within a fractional quantum Hall state of ν = 1/3. We witnessed a pristine AB pattern while observing the interference pattern at a constant filling factor. As we departed from the trajectory of constant filling towards a trajectory of constant density, an evolution of phase jumps emerged within the interference pattern. We showed that the phase accumulated by an interfering quasi-particle may be understood as $2\pi N$, with $N$ being the (continuously varying) number of electrons within the interference loop, and the surge in phase jumps occurring when leaving the constant filling constraint. These measurements open a new platform for studying abelian anyons. Moreover, bilayer graphene offers a plethora of even-denominator FQH states that are expected to realize non-abelian statistics [26,28,29,41,42], which holds great promise for this research direction.

**Author contributions**

J. K. and R. K. improved the quality of the stacks. J. K. and H. D. prepared the stacks. K. W. and T. T. grew the hBN crystals. J. K., A. I., V. B., and R. K. improved the quality of the device. J.K. fabricated the device. A. H. and C. H. developed the measurement circuit and a dilution refrigerator. J.K. performed the measurements. J. K., H. D., A.S., and Y. R. analyzed the measured data. A.S. developed the theoretical aspect. J. K., H. D., A.S., and Y. R. authored the paper with input from all coauthors. Y. R. supervised the overall work done on the project.

**Acknowledgments**

This work was supported by the Quantum Science and Technology Program 2021, by a research grant from the Schwartz Reisman Collaborative Science Program, which is supported by the Gerald Schwartz and Heather Reisman Foundation, by a research grant from the Center for New Scientists at the Weizmann Institute of Science, by grants from the ERC under the European Union's Horizon 2020 research and innovation programme (Grant Agreements LEGOTOP No. 788715 and HQMAT No. 817799), the DFG (CRC/Transregio 183, EI 519/7-1), and by the ISF Quantum Science and Technology (2074/19). Correspondence and requests for materials should be addressed to Y.R. (yuval.ronen@weizmann.ac.il).

**Methods**

**Stack preparation**

The stack used for device fabrications in the current studies is a dual-gated stack with 5 layers. Bilayer graphene flakes are exfoliated mechanically from the bulk graphene crystals utilizing blue tape. To achieve this, the $SiO_2$/Si substrate is cut into small pieces (10 mm × 10 mm), which are then placed on the blue tape and heated at 170 °C ~ 180 °C for 1 minute and 30 seconds on the hot plate. When the tape cools down, these pieces are removed to search for the desired graphene flakes. hBN flakes are obtained with a thin PDMS (tm DGL-30-x4) on $SiO_2$/Si substrate heated at 80 °C for a minute. Due to the fragile nature of hBN, the pieces of $SiO_2$/Si substrate are gently removed from PDMS. To prepare the stack using the dry transfer method, the stamps are placed on a glass slide on which a diamond shaped PDMS working as a cushion for vdW materials is covered with polycarbonate (PC) film held by a Kapton tape. The stamps are placed for approximately 2 hours on a hot plate at 170 °C ~ 180 °C to make PC film adhere to PDMS strongly. The functionality of the PC stamp is first checked on a clean piece of bare $SiO_2$/Si substrate. The transfer stage is heated at 130 °C and all materials (top graphite → top hBN → bilayer graphene → bottom hBN → bottom graphite) get picked up sequentially with the prepared stamp. The stack with a thickness of 48 (32) nm for the top (bottom) hBN and 12 (6) nm for the top (bottom) graphite is used in the device fabrication for the current studies. Finally, the stack is landed at 180 °C on a clean $SiO_2$/Si substrate and left for 20 minutes at 180 °C to melt PC and make PC detached from PDMS. The stack is left for cleaning for 3 to 4 hours in chloroform, and subsequently cleaned with isopropyl alcohol (IPA) and deionized (DI) water. Thermal annealing with ultra-high vacuum (~$10^{-9}$ Torr) is performed at 400 °C for 4 hours to remove partially leftover residues and bubbles. As a last step for the stack preparation, atomic force microscopy (AFM) cleaning is performed to clean and flatten the local area where the device is fabricated.

**Device fabrication**

The standard nanofabrication and lithography methods are employed to fabricate a Fabry Perot interferometer (FPI) device. FPI is fabricated on a highly P-doped Si substrate with a 280 nm $SiO_2$ oxide layer, serving as a metallic and dielectric layer to dope the contact regions at low temperature. The device fabrication starts with forming the align markers for e-beam lithography and bonding pads with Ti 10/Au 60/Pd 20 nm. The device geometry is defined by reactive ion etching (RIE) using polymethyl methacrylate (PMMA) resist as an etch mask. Two types of etching gas are used for etching graphite or hBN: pure $O_2$ for etching graphite, and $O_2$/$CHF_3$ mixture with the volume ratio of 1:10 for etching hBN. After defining the geometry, thermal annealing with ultra-high vacuum (~$10^{-9}$ Torr) is performed at 350 °C for 2 hours to remove the resist residue on the stack. Edge contacts on bilayer graphene are formed by RIE with $O_2$/$CHF_3$ to etch the top hBN layer, and subsequently, evaporation of Cr 2/Pd 13/Au 60 nm with a tilted angle. To form a trench with 40 nm width on the top graphite, RIE with mild $O_2$ plasma condition to reduce the damage on the top hBN is performed using a thinner PPMA resist as an etch mask, thereby cutting the top graphite into eight pieces. Bridges giving access to control on top graphite gates are patterned with PMMA/MMA/PMMA trilayer followed by $O_2$ plasma for 20 seconds and subsequent evaporation of Cr 25/Au 320 nm. Between the fabrication steps, the connectivity between contacts and gates is checked by measuring two-probe resistance in the probe station at room temperature.

**Measurements**

The device is measured in a highly filtered dilution refrigerator at a base temperature $T$ ~ 30 mK using the standard low-frequency lock-in amplifier techniques. SRS 865A lock-in amplifier is used to generate an alternating voltage of 17.7 Hz and measure the voltage difference between two contacts. Putting the load resistance 100 MΩ in series, an alternating voltage with the amplitude of 5 µV (50 µV) is applied to bias an alternating current with the amplitude of 50 pA (500 pA) for fractional (integer) interference measurements. Keithley 2400 is used for applying the voltage on highly P-doped Si substrate to dope the contact region. QDAC (Ultra-low-noise 24-channel DAC) is used to apply and tune the voltage on all graphite gates and two additional air bridges.

**Figures Captions**

**Fig. 1 | Fabry Pèrot Interferometer (FPI) based on bi-layer graphene. a**, False-color SEM image of our FPI. Ohmic contacts are highlighted in yellow. Eight separate regions of top graphite (purple) are formed by etched trenches (pink), connected to air bridges (blue) to apply potentials on each gate separately. Two additional air bridges (green) hover 200 nm above each QPC region to fine-tune individually each saddle point potential. The FPI lithographic area, the area within the interference loop, is 1 μm². Scale bar is 1μm. Inset: top graphite gate etching architecture. **b**, Illustration of an FPI operated at a large perpendicular magnetic field in the QHE. Bias current $I_{SD}$ is applied from source to drain, propagating via edge modes depicted by red lines in the bilayer graphene. The QPCs individual transmission is controlled by either the back gate (BG) or the two air bridges (LBG and RBG). Diagonal resistance $R_D = (V_D^+ - V_D^-)/I_{SD}$ is measured. Base temperature is 30 mK.

**Fig. 2 | Tunable IQHE interference regimes, from CD to AB. a-b, e-f,** Oscillating $\Delta R_D$ as a function of $B$ and $V_{PG}$ at different operating regimes: Interference of the first-LL (LL1) for ν = 1 beneath LG and RG regions, and **(a)** ν = 1 beneath CG region ($V_{CG}$ = 0 V), and **(b)** ν = 2 beneath CG region ($V_{CG}$ = 0.6 V). Interference of the second-LL (LL2) for ν = 2 beneath LG and RG regions, and **(e)** ν = 2 beneath CG region ($V_{CG}$ = -0.1 V), and **(f)** ν = 3 beneath CG region ($V_{CG}$ = 0.8 V). **c-d, g-h**, 2D-FFT corresponding to **(a-b)**, **(e-f)**, respectively. Insets: LLs and Fermi level, $E_F$, beneath the CG region. Red dots describe the interfering edge channel in the CG region. **i**, Measured $\frac{\Phi_0}{\Delta B}$ as a function of $V_{CG}$ for the two LL1 (blue) and LL2 (green). Negative values point to the CD regime. Labels of initial and final correlate to the labels on **(a)-(h)**. The blue (red) background color describes the AB (CD) regime. $V_{BG}$ is set to 0.5 V for LL1, and 1.05 V for LL2 interference.

**Fig. 3 | AB interference at a 1/3 fractional filling. a**, $R_{xx}$ fan diagram performed on the right side of the FPI. **b-c**, oscillating $\Delta R_D$ in the plane of $B$- $V_{PG}$ for: **(b)** constant bulk filling ν = 1/3 beneath LG, RG, and CG regions, showing fractional AB interference. **(c)** bulk filling is $\nu = \frac{1}{3}$ beneath the LG and RG regions, and ν = 1 beneath CG region, showing CD interference. **d-e**, 2D-FFT corresponding to **(b)-(c)**, respectively. Insets: Schematic of the edge modes in the two configurations, **(d)** quasi-particle tunneling, and **(e)** electron tunneling, with dashed lines. **f**, $\frac{\Phi_0}{\Delta B}$ obtained from 2D-FFT and $R_{xx}$ obtained from Fig. 3a line cut at 10.5 T as a function of $V_{CG}$. The blue (red) background color describes the AB (CD) regimes. Labels of initial and final correlate to the labels on **(b)-(e)**. $V_{BG}$ is set to 0.5 V for all measurements in Fig. 3.

**Fig. 4 | Tunability between constant filling and constant density. a**, $R_{xx}$ fan diagram, in the $B$-$V_{RG}$ plane, zoom-in to ν = 1/3 region. Three white solid lines (①-③) describe the trajectories which the voltages on all three top gates (L/RG and CG) follow while $B$ is varied. **b-d**, Oscillating $\Delta R_D$ in the $B$-$V_{PG}$ plane for: **(b)** constant filling ν = 1/3 beneath LG, CG, and RG regions (trajectory ①). **(c)** intermediate regime between constant filling and density (trajectory ②). **(d)** constant density where the voltages applied on LG, CG, and RG are fixed (trajectory ③). **e**, Summary of $\frac{\partial \theta}{\partial B}$, and $\chi_{fit}$ (see Fig. S13), extracted from 2D interference patterns at altered trajectories in (a), $\alpha = \frac{dB}{dV_{CG}}$. The blue background color describes the AB-dominated regime, and phase slips emerge to the left and right. Inset: qualitative schematic for the density of state of hole-like and particle-like states as a function of $E$, and position of Fermi level $E_F$. $V_{BG}$ is set to 0.5 V for all measurements in Fig. 4.

**Figures**

**Fig. 1**

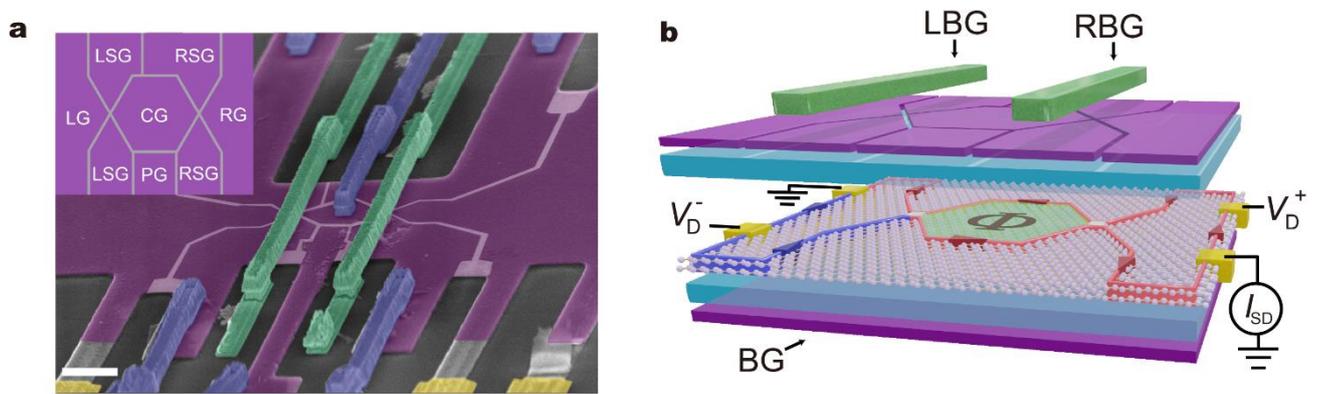

**Fig. 2**

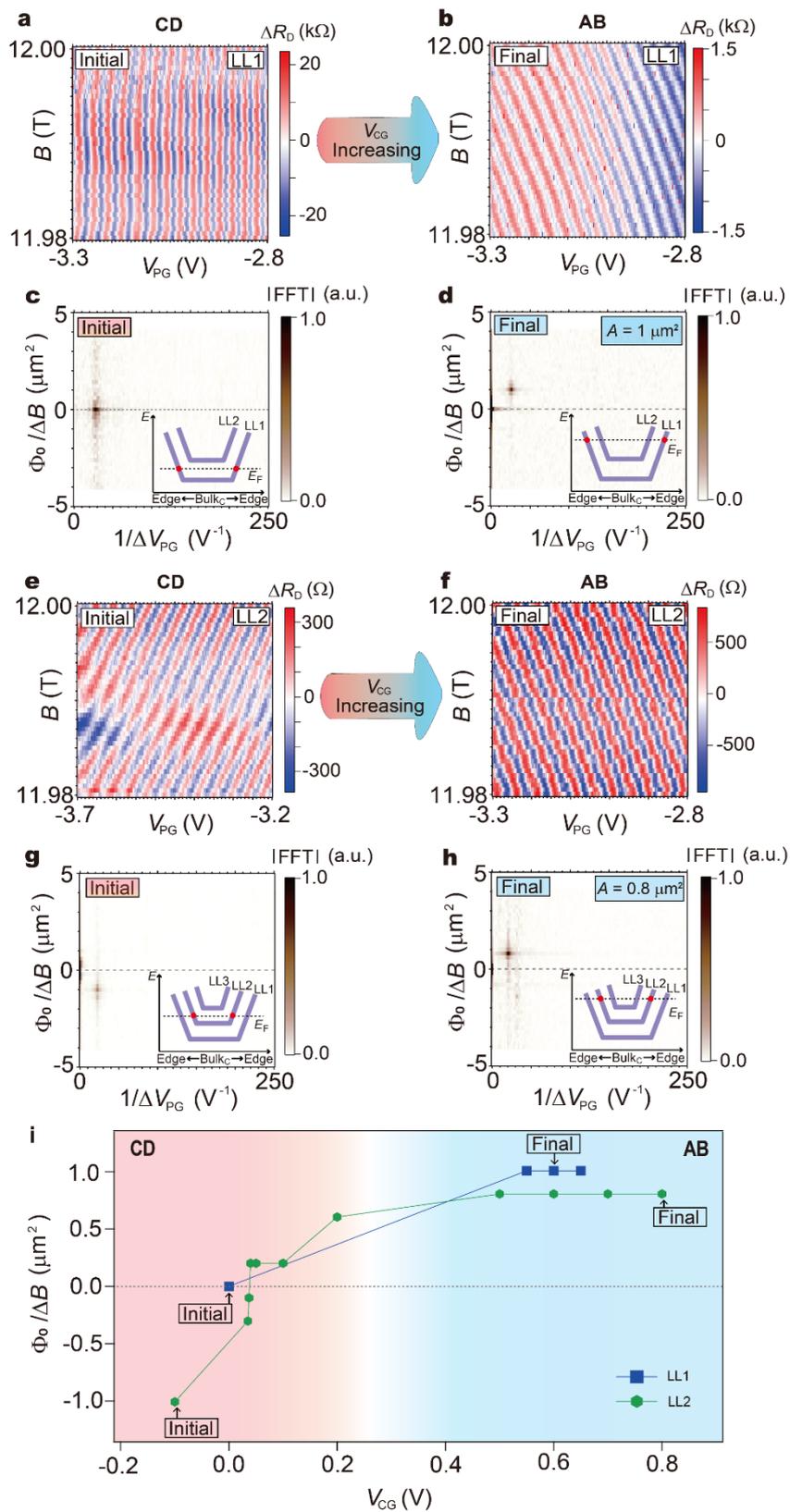

**Fig. 3**

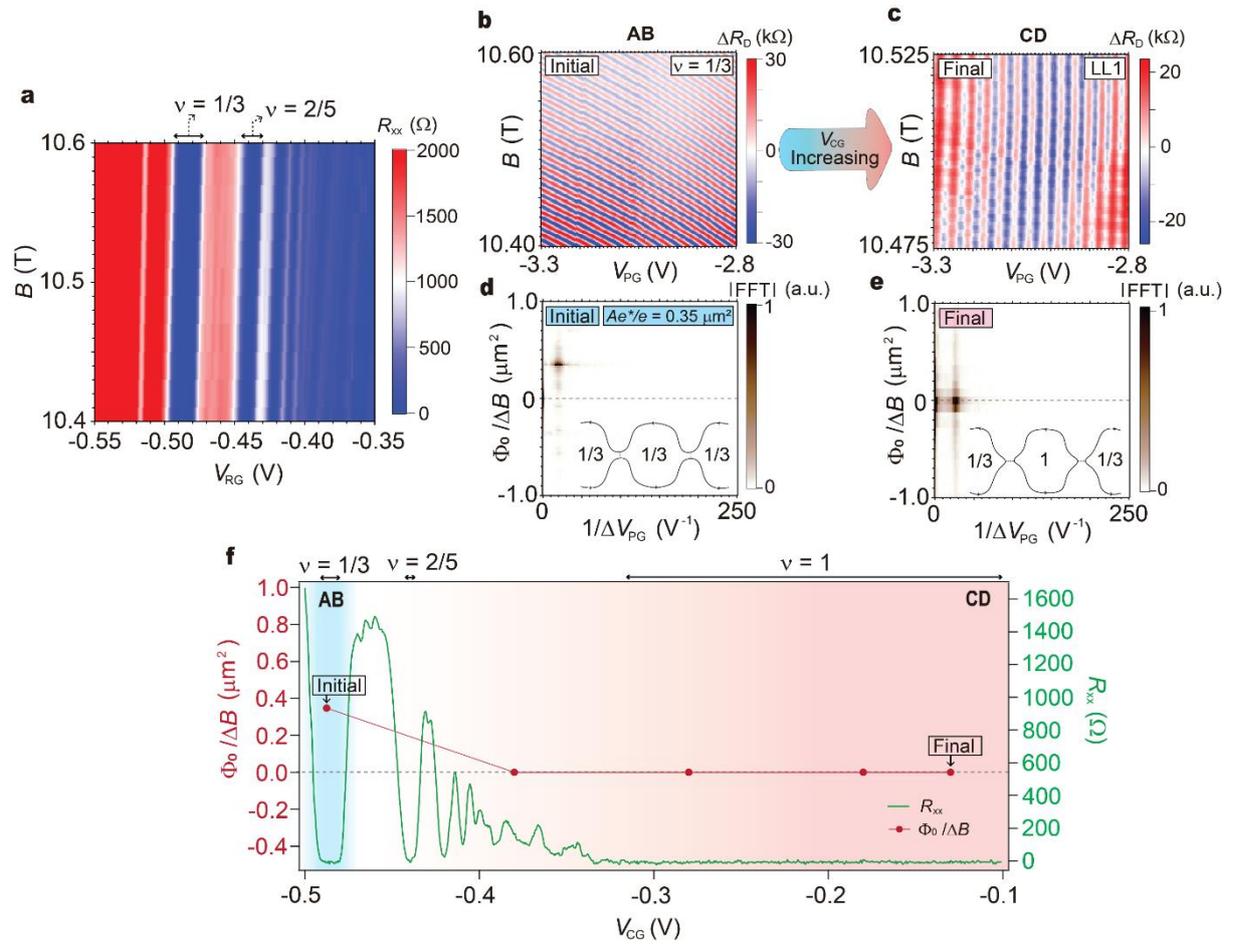

**Fig. 4**

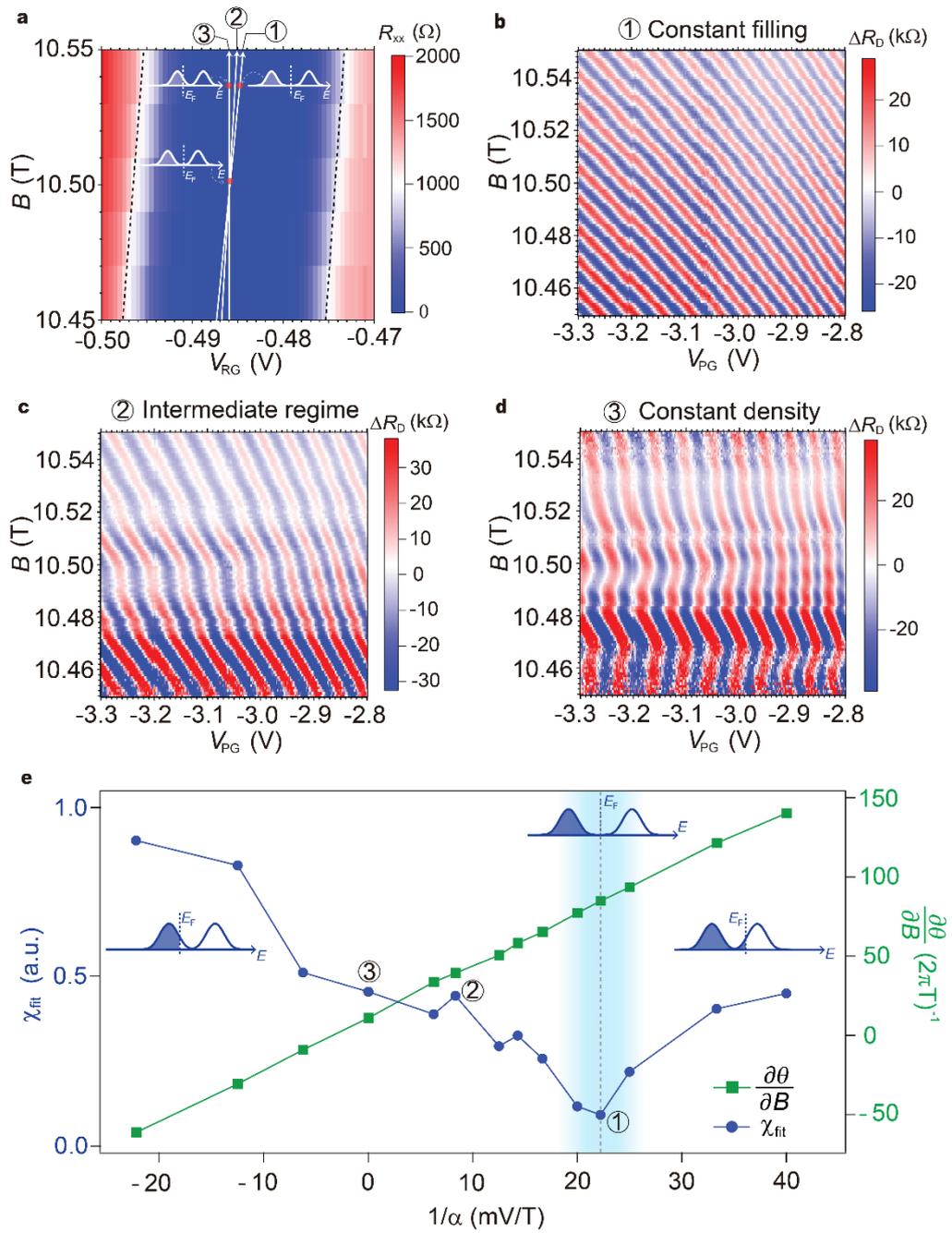

# Supplementary information

**Aharonov-Bohm interference and the evolution of phase jumps in fractional quantum Hall Fabry-Perot interferometers based on bi-layer graphene**

Jehyun Kim[1], Himanshu Dev[1], Ravi Kumar[1], Alexey Ilin[1], André Haug[1], Vishal Bhardwaj[1], Changki Hong[1], Kenji Watanabe[2], Takashi Taniguchi[3], Ady Stern[1], and Yuval Ronen[1]*

[1] *Department of Condensed Matter Physics, Weizmann Institute of Science, Rehovot, Israel*

[2] *Research Center for Functional Materials, National Institute for Materials Science, Tsukuba, Japan*

[3] *International Center for Materials Nano architectonics, National Institute for Materials Science, Tsukuba, Japan*

\* *Corresponding author*: yuval.ronen@weizmann.ac.il

**Supplementary Section 1: Device characterization**

**Supplementary Section 2: QPCs characterization**

**Supplementary Section 3: $V_{CG}$ dependence of the interference at IQHE**

**Supplementary Section 4: $V_{CG}$ dependence of the interference at FQHE**

**Supplementary Section 5: Gate tunable AB interference**

**Supplementary Section 6: Trajectory dependence of 1/3 interference**

**Supplementary Section 7: Constant density/filling in the reservoirs/interference area**

**Supplementary Section 8: Phase slips**

**Supplementary Section 9: Extracting the values of phase slips from 1D-FFT**

**Supplementary Section 10: AB interference at ν = - 1/3 fractional filling**

**Supplementary Section 1: Device characterization**

Fig. S1a shows the optical microscope image for one of our FPI devices, fabricated at the same batch. One of the widely used methods to characterize the quality of a device is measuring its mobility. Devices with high mobility are required to observe fractional quantum Hall states, thus, we perform the mobility analysis on another Hallbar device fabricated in a similar condition, by measuring the four-probe resistance $R_{xx}$ as a function of $V_{BG}$ while applying 10 nA at 30 mK (B = 0 T). The measured $R_{xx}$ is fitted with the equation as follows [1],

$$R_{xx} = \frac{L}{We\mu\sqrt{n_0^2+(\frac{C_{BG}(V_{BG}-V_{DP})}{e})^2}} \tag{S1}$$

where $n_0$ is the intrinsic carrier concentration induced by the charged impurity, $C_{BG}$ is the capacitance per unit area of the bottom graphite gate, $V_{BP}$ is the voltage corresponding the charge neutrality point, and $L$, $W$, $\mu$, and $e$ are the length between the two probe contacts, width of the channel, mobility, and electron charge, respectively. We extract ~240,000 cm²/Vs mobility, high enough to observe the fractional quantum Hall states, and $n_0 = 1.8 \times 10^9$ cm$^{-2}$, reflecting the low charged impurity density in our FPI devices [1,2].

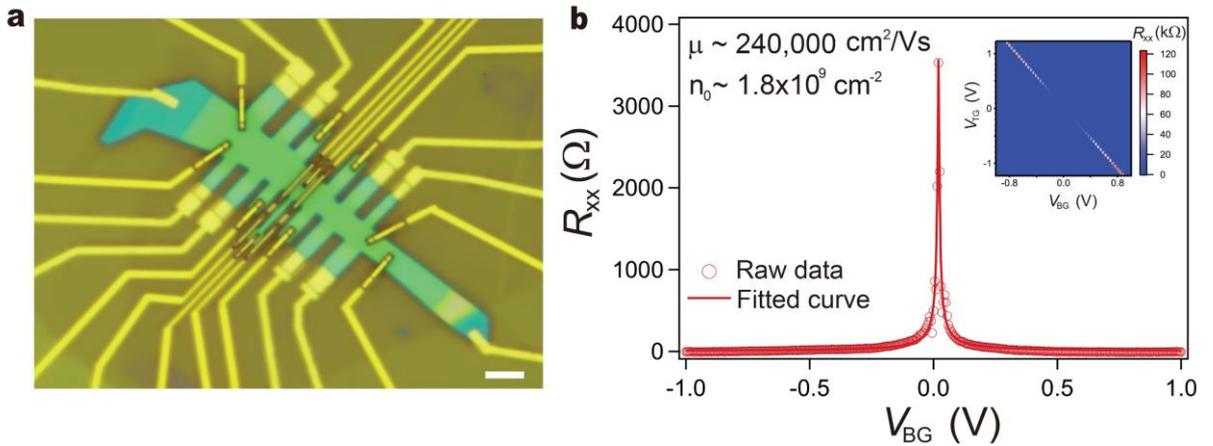

**Fig. S1 | Mobility analysis. a**, Optical microscope image of a bilayer graphene-based FPI. Scale bar indicates 5μm. **b**, Four probe resistance $R_{xx}$ measured as a function of $V_{BG}$ while

applying 10 nA at 30 mK under 0 T. All top graphite gates are set to 0 V. Inset: $R_{xx}$ measured in $V_{TG}$-$V_{BG}$ plane where displacement field is generated in a bilayer graphene.

**Supplementary Section 2: QPCs characterization**

In this section, we describe the characterization of the left and right QPC (L/RQPC) in both IQHE and FQHE. Two air bridges (L/RBG) are employed to control the transmission of the QPCs' separately. Forming each of the QPCs (LQPC or RQPC) separately, we can tune the voltage applied on L/RBG to control the density of each QPC region, thereby tunning the QPCs' transmission given by $t_{\text{L or R}} = G_D \frac{h}{e^2}$ where the diagonal conductance $G_D = \frac{1}{R_D}$.

To form LQPC (RQPC), the filling factors are set to 0 beneath LSG (RSG) and PG regions. Fig. S2a and S2b show LQPC (RQPC) characterization measured with $G_D$ as a function of $V_{LBG}$ ($V_{RBG}$) in the condition where ν beneath LR/G, CG, and RSG (LSG) is set to ν=1, for Fig. S2a, and ν=2 for Fig. S2b while applying 500 pA of $I_{SD}$ under 12 T, clearly showing the QPC partitioning probed by step-like behavior with the plateaus indicating that IQHE edges are fully transmitted across the QPC. In the measurements in Fig. 2 of the main text, $t_L$ and $t_R$ are set to ~0.5 for the interference of LL1, and ~1.5 for the interference of LL2, respectively.

Fig. S2c and S2d also show similar LQPC (RQPC) characterization measured at the condition where ν beneath LR/G, CG, and RSG (LSG) is set to 1/3 while applying 50 pA of $I_{SD}$ under 12 T for Fig. S2c and 10.5 T for Fig. S2d, respectively. One distinct feature which is not observed in Fig. S2a-b is many peaks emerging on the plateau, which may be attributed to FQHE edges vulnerable to the charge disorder near QPCs saddle point and etched trenches. In the measurements in Fig. 3 and 4 of the main text, $t_L$ and $t_R$ are set to ~0.75 and ~0.95, respectively.

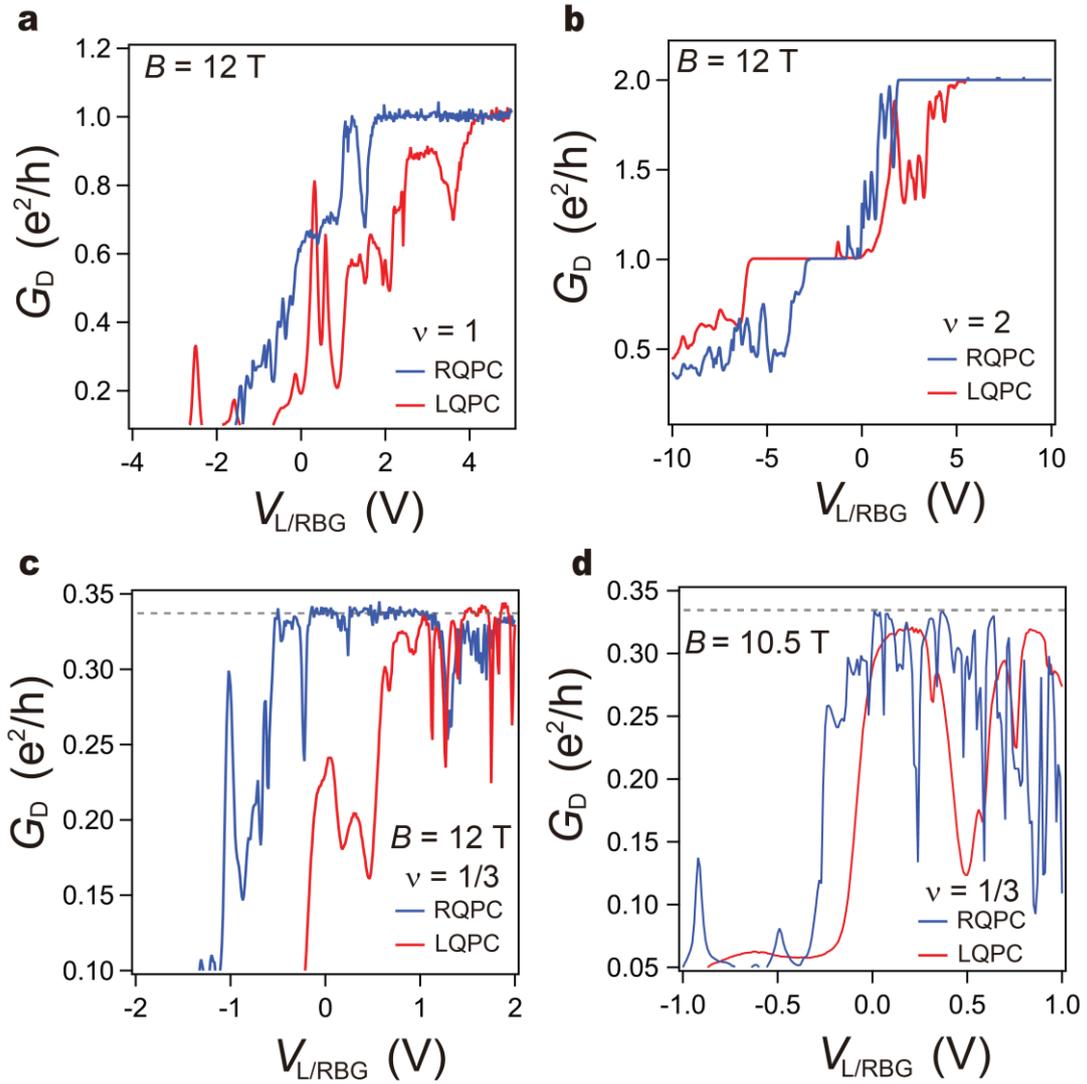

**Fig. S2 | QPCs characterization in the IQHE and FQHE. a-d,** Diagonal conductance $G_D$ for LQPC (red) and RQPC (blue) as a function of either $V_{LBG}$ or $V_{RBG}$ at different filling factors. (a) $\nu = 1$ at 12 T. (b) $\nu = 2$ at 12 T. (c) $\nu = 1/3$ at 12 T. (d) $\nu = 1/3$ at 10.5 T. Note a single ground is employed in these measurements.

**Supplementary Section 3: $V_{CG}$ dependence of the interference at IQHE**

We present all the 2D pajama plots and the corresponding 2D-FFT results used for analyzing $V_{CG}$ dependence of the LL1 and LL2 interference shown in Fig. 2 of the main text. Fig. S3 (S5) and S4 (S6) show 2D pajama of $R_D$ as a function of $B$ and $V_{PG}$ measured at different $V_{CG}$, and the corresponding 2D-FFT, relevant to the interference of LL1 (LL2). Note that 2D pajama at $V_{CG} = 0.2$ V in Fig. S3b shows the intermediate regime between CD and AB regimes where the phase of the oscillating $R_D$ is not stable.

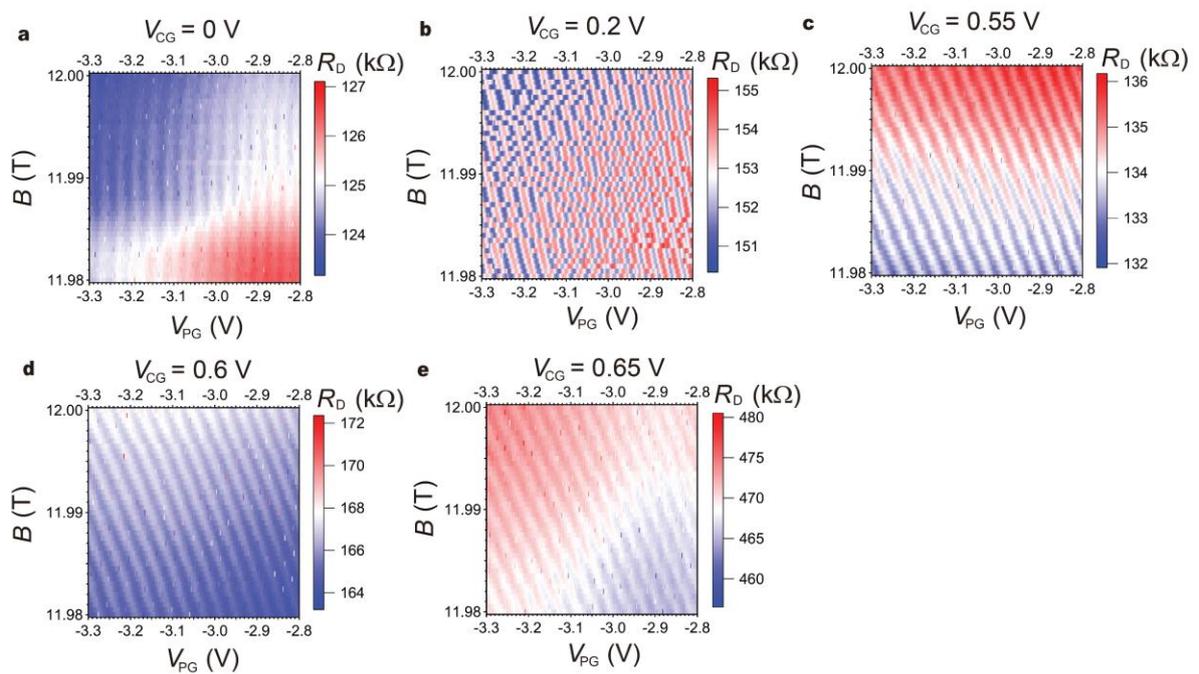

**Fig. S3 | $V_{CG}$ dependence of LL1 interference. a-e**, Oscillating $R_D$ as a function of $B$ and $V_{PG}$ at different $V_{CG}$.

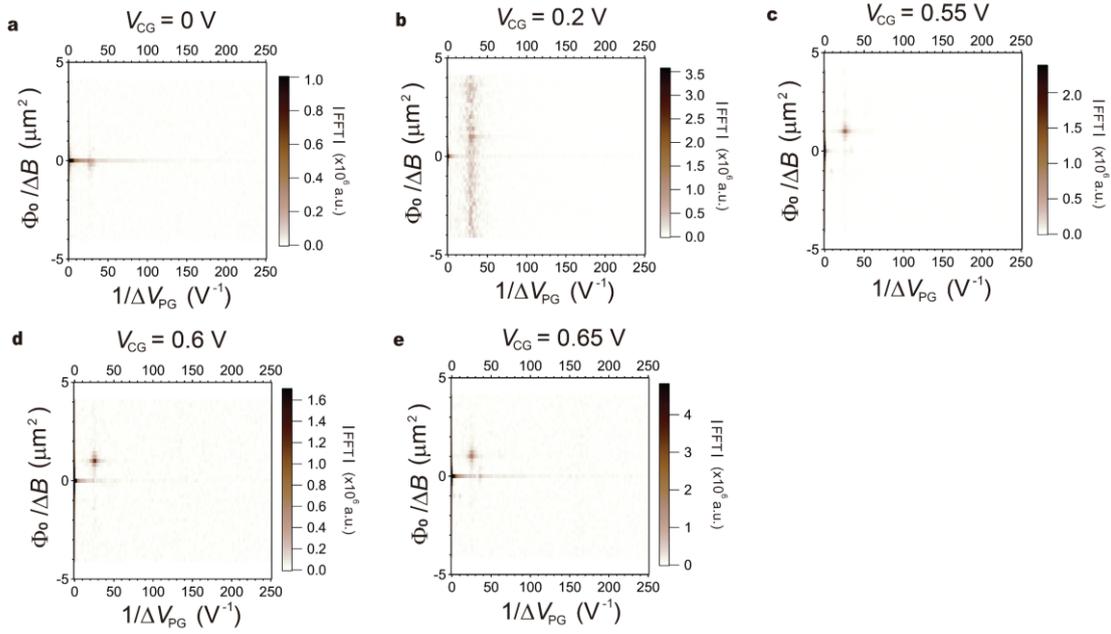

**Fig. S4 | 2D-FFT analysis on $V_{CG}$ dependence of LL1 interference. a-e**, 2D-FFT corresponding to Fig. S3 a-e, respectively.

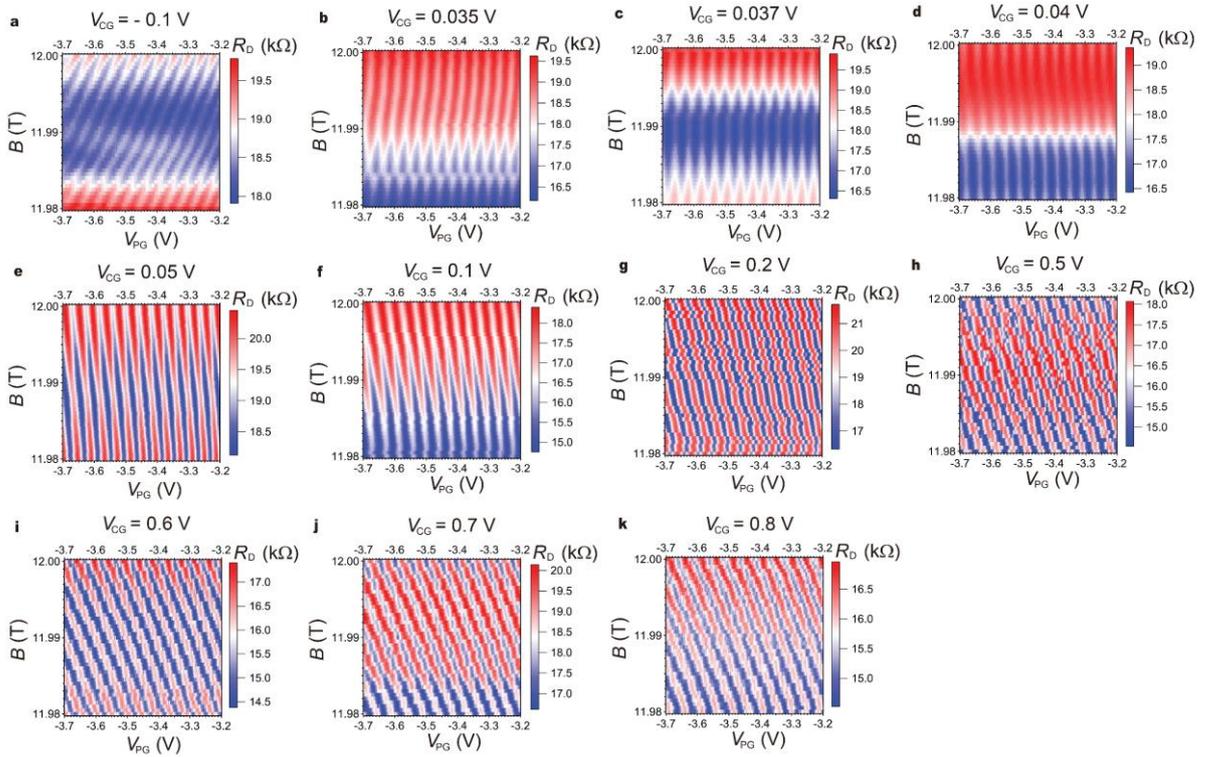

**Fig. S5 | $V_{CG}$ dependence of LL2 interference. a-k**, Oscillating $R_D$ as a function of $B$ and $V_{PG}$ at different $V_{CG}$.

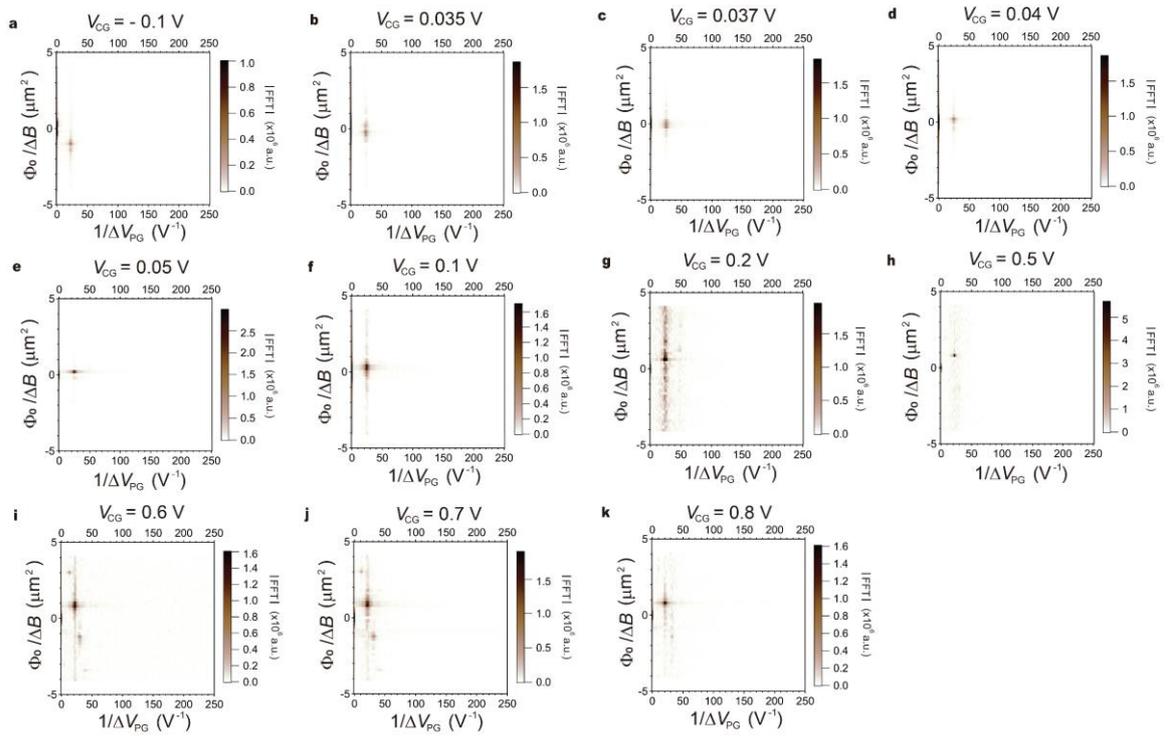

**Fig. S6 | 2D-FFT analysis on $V_{CG}$ dependence of LL2 interference. a-k**, 2D-FFT corresponding to Fig. S5 a-k, respectively.

**Supplementary Section 4: $V_{CG}$ dependence of the interference at FQHE**

We present all 2D pajama plots and the corresponding 2D-FFT results used for analyzing $V_{CG}$ dependence of the interference at a 1/3 fractional filling shown in Fig. 3 of the main text. Fig. S7 and S8 show 2D pajama of $R_D$ as a function of $B$ and $V_{PG}$ measured at different $V_{CG}$, and the corresponding 2D-FFT, clearly showing the crossover from AB to CD-dominated oscillations. As $V_{CG}$ is increased further, at $V_{CG}$ = - 0.05 V (Fig. S7f), 2D oscillations with negative slopes of the constant phase lines start to emerge in the intermediate regime between CD and AB regimes where the phase of the oscillating $R_D$ is not stable, implying the interfering edge is not fully developed, similar to the result shown in Fig. S3b. The same measurements are performed at a higher magnetic field near 11.985 T, shown in Fig. S9 for 2D pajama plots and Fig. S10 for the corresponding 2D-FFT, showing similar behavior (AB → CD → AB-dominated oscillations) with increasing $V_{CG}$. Note that Fig. S9a is measured at a constant density condition where non-linearity in the constant phase lines emerges.

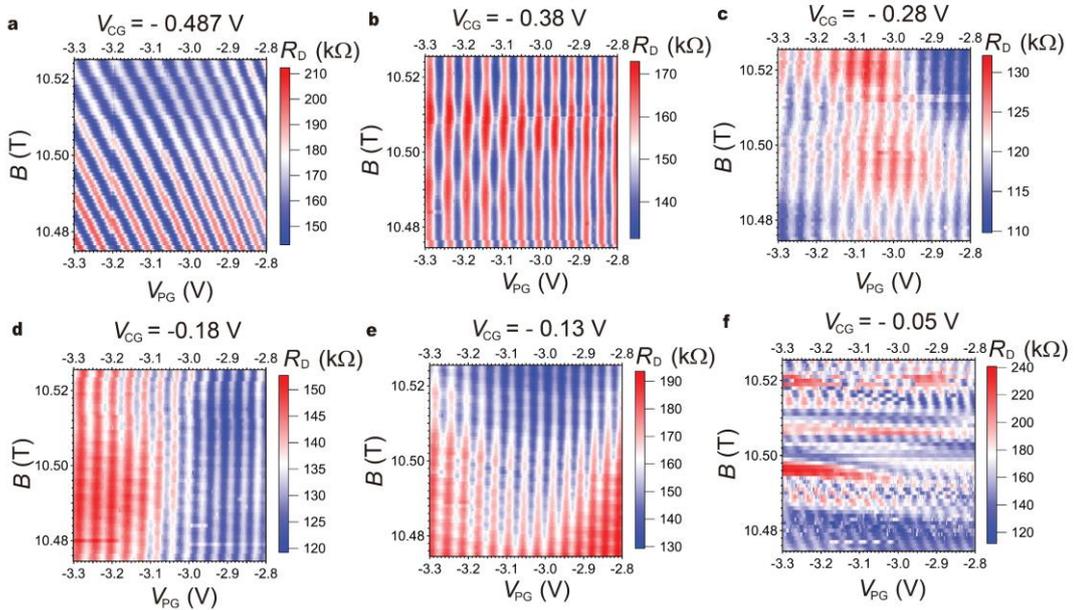

**Fig. S7 | $V_{CG}$ dependence of an interference at a 1/3 fractional filling near 10.5 T. a-f**, Oscillating $R_D$ as a function of $B$ and $V_{PG}$ at different $V_{CG}$.

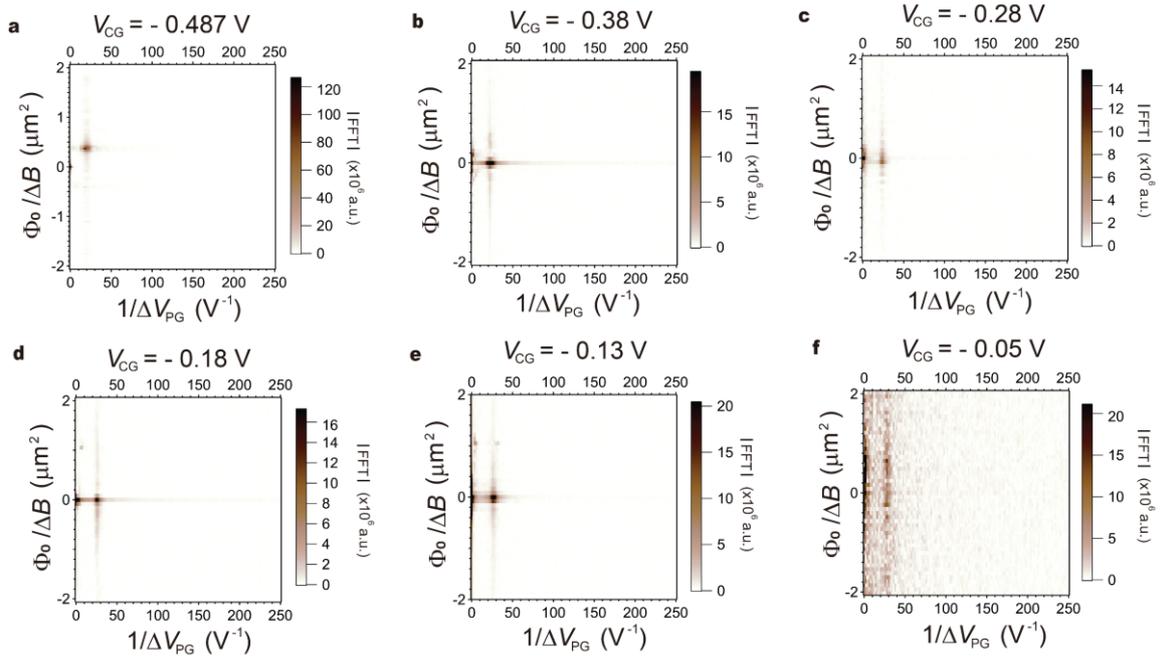

**Fig. S8 | 2D-FFT analysis on $V_{CG}$ dependence of an interference at a 1/3 fractional filling near 10.5 T. a-f**, 2D-FFT corresponding to Fig. S7a-f, respectively.

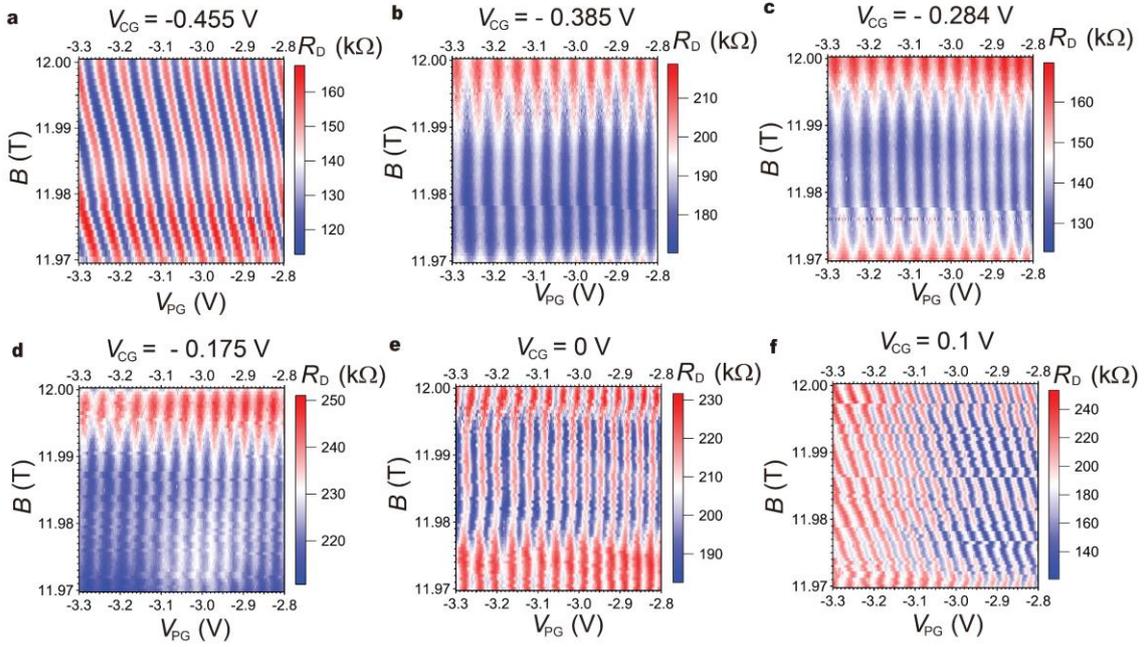

**Fig. S9 | $V_{CG}$ dependence of an interference at a 1/3 fractional filling near 11.985 T. a-f**, Oscillating $R_D$ as a function of $B$ and $V_{PG}$ at different $V_{CG}$.

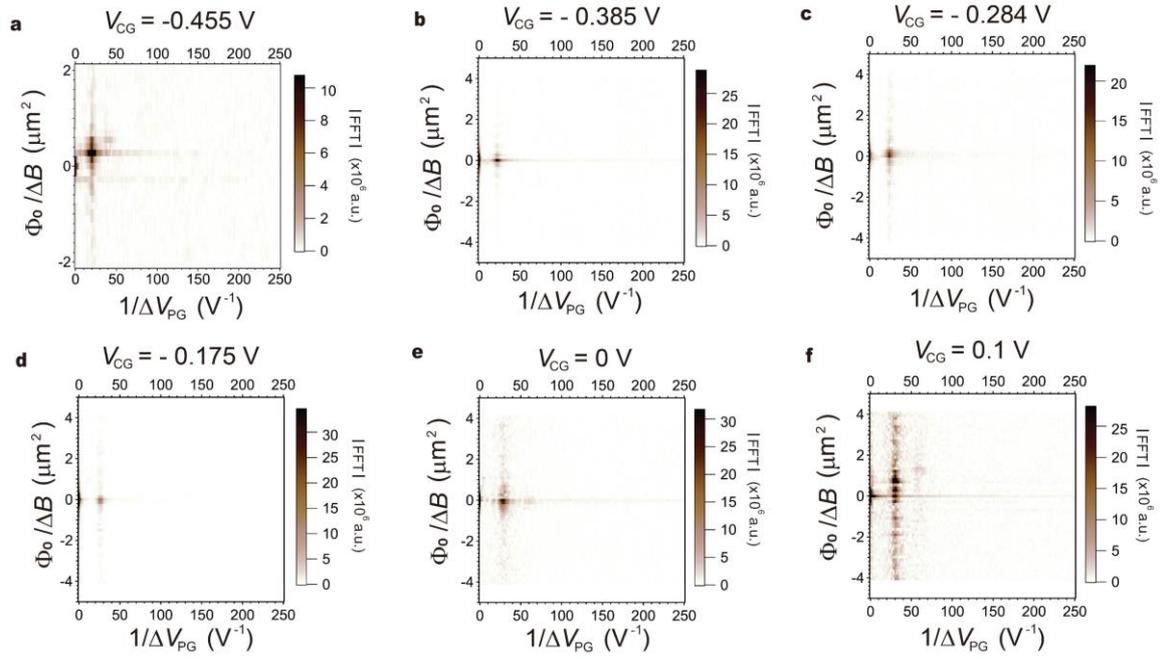

**Fig. S10 | 2D-FFT analysis on $V_{CG}$ dependence of an interference at a 1/3 fractional filling near 11.985 T. a-f**, 2D-FFT corresponding to Fig. S9a-f, respectively.

**Supplementary Section 5: Gate tunable AB interference**

In this section, we study the electrostatics of the plunger gate by analyzing the $V_{PG}$ dependence of AB oscillation frequency. In a similar measurement condition used in Fig. 3b, as shown in Fig. S11c, AB oscillations are measured as a function of $V_{PG}$ plotted with $R_D$ at ν = 1/3 beneath LG, CG, and RG regions under 12 T, clearly showing the oscillating frequency depending on $V_{PG}$, which is the same behavior as that reported in the previous studies in mono and bi-layer graphene-based FPI [3–5]. From eq. 1 in the main text, periodicity in $V_{PG}$, $\Delta V_{PG}$, can be derived as $\Delta V_{PG} = \frac{e}{Be^*}\frac{\Phi_0}{dA/dV_{PG}}$, where $\frac{dA}{dV_{PG}}$ is the lever arm indicating the effect of $V_{PG}$ on the interfering area $A$ [6]. As long as $\frac{dA}{dV_{PG}}$ is constant value at different electrostatic environment near the interference loop, $\Delta V_{PG}$ can give the information of the interfering charge. However, in graphene-based FPI, $\frac{dA}{dV_{PG}}$ is a function of $V_{PG}$, as shown in Fig. S11c, which implies that the spatial distance between the trajectory of the QH edge channel and the PG can be affected by ν beneath PG region. Moreover, it is reported that $\frac{dA}{dV_{PG}}$ varies with the different ν beneath CG region [3]. Therefore, this complexity in $\frac{dA}{dV_{PG}}$ depending on ν beneath both CG and PG regions makes it harder to extract the correct interfering charge from the measured $\Delta V_{PG}$. We note that the visibility of the oscillations, defined as $\frac{R_{D,max}-R_{D,min}}{R_{D,max}+R_{D,min}}$ where $R_{D,max}$ and $R_{D,min}$ are maximum and minimum values, respectively, is rather high, around 35%, indicating the high fractional edge mode velocity in our FPI [3,4,6].

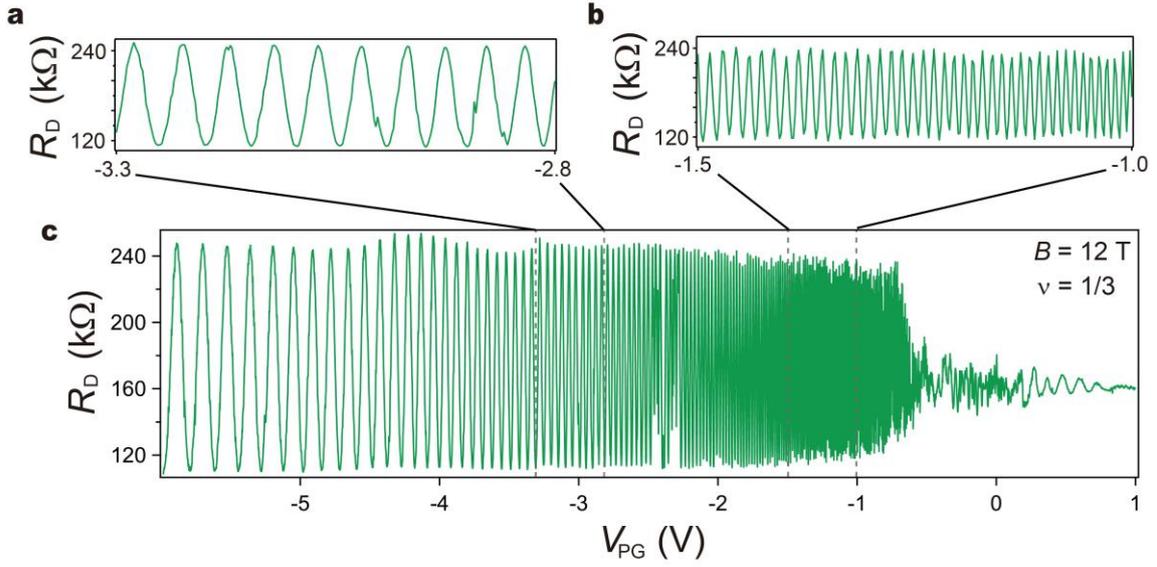

**Fig. S11 | Plunger gate dependence of 1/3 filling. a-c,** Diagonal resistance $R_D$ oscillations as a function of plunger gate voltage ($V_{PG}$) measured for $\nu = 1/3$ at a fixed magnetic field 12 T, (**a**) $R_D$ oscillations on partial $V_{PG}$ range from -3.3 to -2.8 V. (**b**) $R_D$ oscillations on partial $V_{PG}$ range from -1.5 to -1.0 V, showing higher oscillating frequency as $V_{PG}$ is increased. (**c**) $R_D$ oscillations in a whole range of $V_{PG}$, showing the suppression of the oscillations amplitude above -0.6 V of $V_{PG}$ corresponding to the charge neutrality point. $I_{SD}$=50 pA is applied with QPCs transmission of $t_L \sim 0.75$ and $t_R \sim 0.95$ under 12 T. $V_{BG}$ is set to 0.5 V.

**Supplementary Section 6: Trajectory dependence of 1/3 interference**

We present all the 2D pajama plots and the corresponding 1D-FFT results used for analyzing trajectory dependence of 1/3 interference shown in Fig. 4e of the main text. Fig. S12 and S13 show the 2D pajama of $R_D$ as a function of $B$ and $V_{PG}$ measured at different trajectories, and the corresponding 1D-FFT, respectively. 1D-FFT analysis is performed on the results in Fig. S12a-n to obtain the phase evolution $\frac{\theta}{2\pi}$ as a function of $B$, plotted in Fig. S13a-n. The detailed method for 1D-FFT is explained in the supplementary section 9. To quantify the linearity of the constant phase lines in the 2D oscillations, we employ a least-squares fitting on 1D-FFT results with the linear function, yielding $\frac{\partial \theta}{\partial B}$ and $\chi_{fit}$ where $\frac{\partial \theta}{\partial B}$ is the slope obtained from the least-squares fits (black dashed line) in the unit of $(2\pi T)^{-1}$, and $\chi_{fit}$ is the standard deviation calculated in the least-squares fitting process. The $B/V_{CG}$ slope $\alpha = 45$ T/V, in which $\chi_{fit}$ is in its minimum corresponds to the constant filling condition where pristine AB oscillations are observed.

The linear behavior of $\frac{\partial \theta}{\partial B}$ with $\frac{1}{\alpha}$, shown in Fig. 4e of the main text, can provide the simple relation between the phase and the number of electrons in the interference loop, as described in the main text. The slope $\alpha$ that defines the trajectory can be rewritten:

$$\frac{1}{\alpha} = \frac{\partial V_{CG}}{\partial B} = \frac{1}{C}\frac{\partial Q}{\partial B} = \frac{e}{C}\frac{\partial N}{\partial B} \tag{S2}$$

where $Q$ and $N$ are total charge and the number of electrons in the interference loop, respectively, and $C$ is the capacitance between CG and the bulk of the interference region. Using eq. S2, the slope, $\beta$, obtained from the least-squares fits of the result in Fig. 4e of the main text is then:

$$\beta = \frac{\frac{1}{2\pi}\frac{\partial \theta}{\partial B}}{\frac{e\partial N}{C\partial B}} = \frac{C}{2\pi e}\frac{\partial \theta}{\partial N} \ . \tag{S3}$$

From eq. S3, the phase difference can be expressed as follows:

$$\partial\theta = \partial N\, 2\pi(\beta\frac{e}{C}) \tag{S4}$$

To calculate the capacitance, C, including the quantum capacitance, we use the Streda formula. For a 1/3 filling, $1/3 = n_e/n_\phi = \Phi_0 \cdot \frac{C \cdot V_{CG}}{e \cdot A \cdot B}$ leading to $C = \frac{e^2}{3 \cdot h} \cdot A \cdot \frac{dB}{dV_{CG}}$. Taking $A = 1.05 \mu m^2$, 3 times the area measured from the constant filling trajectory and $\frac{dB}{dV_{CG}} = 45 T/V$, we calculate $C = 6.08 \times 10^{-16}$ F, somewhat lower than the geometric capacitance $C \sim 6.6 \times 10^{-16}$ F. The value of $\frac{dB}{dV_{CG}}$ is measured both from the average of the black dashed lines in Fig. 4a as well as from the min-error fit in Fig. 4e.

Using the values of $C \sim 6.08 \times 10^{-16}$ F, $e \sim 1.6 \times 10^{-19}$ C, and $\beta \sim 3500$ V$^{-1}$, we obtain $\frac{\partial\theta}{\partial N} = 2\pi(0.92)$, yielding $\theta \approx 0.92 \cdot (2\pi N)$, mentioned in the main body of the paper.

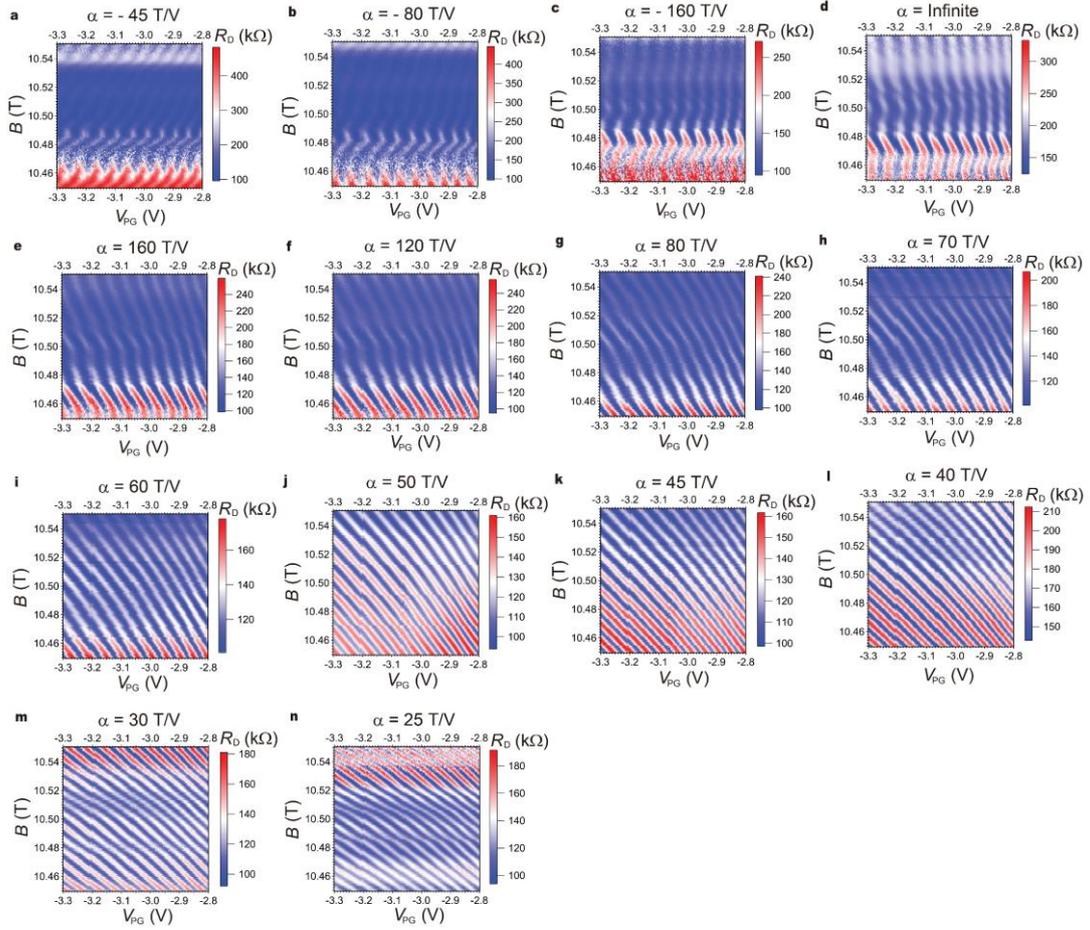

**Fig. S12 | Trajectory dependence of an interference at a 1/3 fractional filling. a-n**, Oscillating $R_D$ as a function of $B$ and $V_{PG}$ at different trajectories. $\alpha$ written above every single image means the trajectory direction where three top gates (L/RG and CG) and $B$ change together.

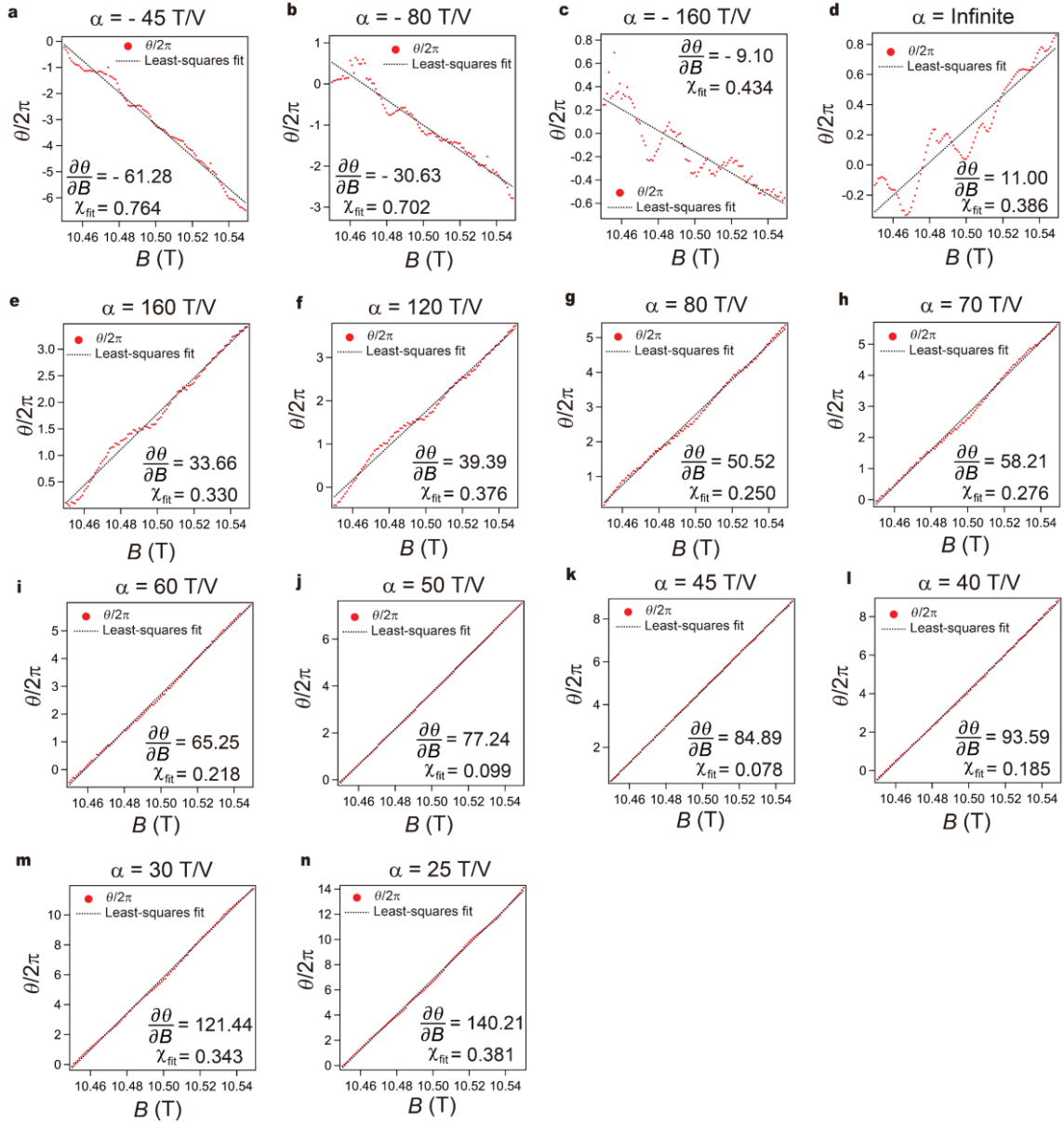

**Fig. S13 | 1D-FFT of trajectory dependence of 1/3 interference. a-n**, Scaled phase difference θ/2π extracted from 1D-FFT of Fig. S12a-n, and corrpesponding least-squares fits, as a function of *B*. Inset: Fitting values $\frac{\partial \theta}{\partial B}$ and $\chi_{fit}$ are presented.

**Supplementary Section 7: Constant density/filling in the reservoirs/interference area**

As an extension of the measurements in Fig. 4 of the main text, we perform a measurement aimed to confirm the absence of the relative contributions to $\theta$ by quasi-particles in the L/RG and CG regions. Fig. S14a shows the 2D oscillations measured at the condition where L/RG regions are in constant density, and CG region is in constant filling, showing pure AB oscillations similar to the result in Fig. 4b of the main text. This result means that the change in $N_{qp}$ in the bulk beneath L/RG regions does not distort the linear lines of the constant phase in AB oscillations. On the other hand, as shown in Fig. S14b, when L/RG regions are in constant filling, and CG region is in constant density, the 2D oscillations show the non-linearity in the lines of the constant phase, resulting in the same behavior as that shown in Fig. 4d of the main text. Therefore, from these control experiments, we can demonstrate that $\Delta N_{qp}$ in eq. 1 of the main text is the number of quasiparticles localized in bulk beneath CG region rather than L/RG regions, and the non-linearity in the lines of the constant phase results from the effect of quasiparticles within the interference loop.

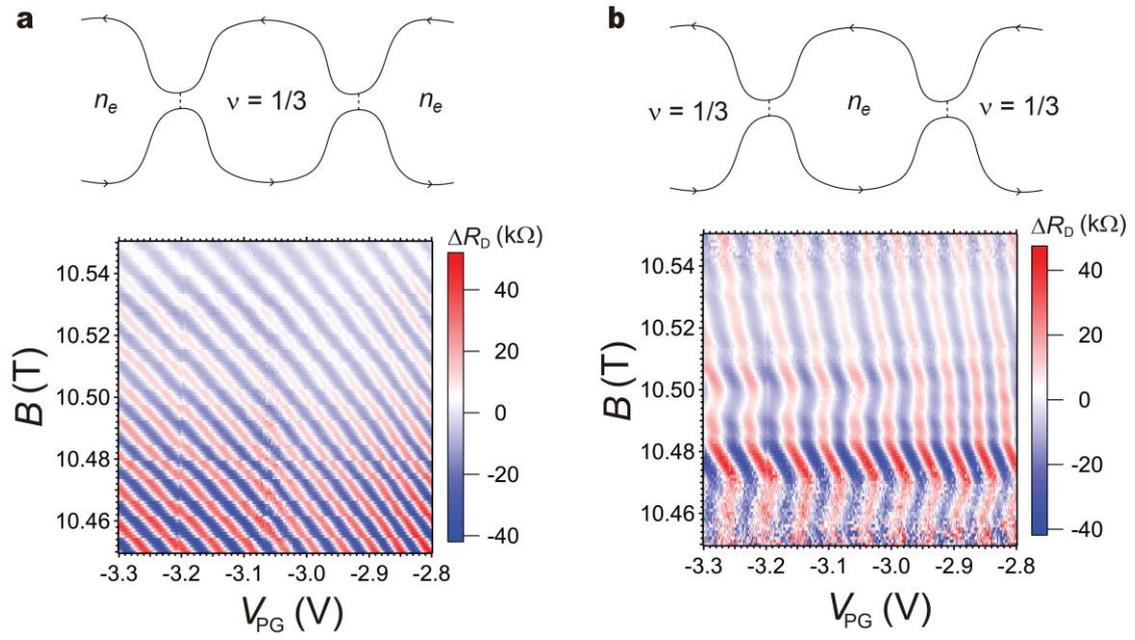

**Fig. S14 | Constant density/filling in the reservoirs/interference area. a-b**, Oscillating $\Delta R_D$ as a function of $B$ and $V_{PG}$ at different measurement conditions. (a) L/RG regions are in constant density, and CG region is in constant filling. (b) L/RG regions are in constant filling, and CG region is in constant density. $n_e$ in the upper schematic of edge channels decribes constant density.

**Supplementary Section 8: Phase slips**

In this section, we present the observed phase slips at ν = 1/3 interfering edge mode. We operate the FPI at a higher $B$ = 12T to increase the 1/3 energy gap $\Delta_{1/3}$, expecting a bigger $\Delta B_C$ where an increased number of phase slips can occur in a wider continuous AB regime. In these measurement conditions, we observed a large number of phase slips (Fig. S15a). We concentrate on three phase slips occurring at 11.99 T, 11.978 T, and 11.961 T, indicated by gray dashed arrows in Fig. S15b. To extract the quantitative value of the phase slips, 1D-FFT along the horizontal lines where phase slips occur is performed. Subsequently, AB phase contribution is subtracted from the total phase difference $\theta$, as shown in Fig. S15c. The values of the phase slips, $\Delta\theta/2\pi$, are calculated by taking a difference between the average values on two adjacent plateaus depicted as dashed lines in Fig. S15c, returning the values of $\Delta\theta/2\pi$ = -0.14, -0.29, and -0.34 deviating from the value of -1/3 predicted in the theory. This deviation can be understood by employing the bulk-to-edge coupling in $\Delta\theta/2\pi$, which can be modified at ν = 1/3 interference as follows [7,8]:

$$\frac{\Delta\theta}{2\pi} = -\frac{\theta_{\text{anyon}}}{2\pi} + \frac{3K_{\text{IL}}}{K_{\text{I}}}\left(\frac{e^*}{e}\right)^2 \tag{S5}$$

, where $K_{\text{IL}}$ represents the coupling of the bulk-to-edge channel, and $K_{\text{I}}$ is the edge stiffness describing the energy cost to vary the interfering area $A$. According to eq. S5, the absolute value of the phase slips can decrease when non-negligible $K_{\text{IL}}$ exists, explaining $\Delta\theta/2\pi$ = -0.14 and -0.29 at 11.99 T and 11.978 T, respectively. In addition, the behavior of different $\Delta\theta/2\pi$ may be attributed to the fact that $\frac{K_{\text{IL}}}{K_{\text{I}}}$ can be affected by $\Delta N_{\text{qp}}$. Another important point is that discrete phase slips occur along the horizontal lines, without depending on $V_{\text{PG}}$ (at these ranges), different then the reported previous work where these phase slips occur along the lines with the positive slope in $B$-$V_{PG}$ plane [8,9], implying that the main reason for $\Delta N_{qp}$ is the change in the degeneracy of the bulk within the interference loop while changing $B$.

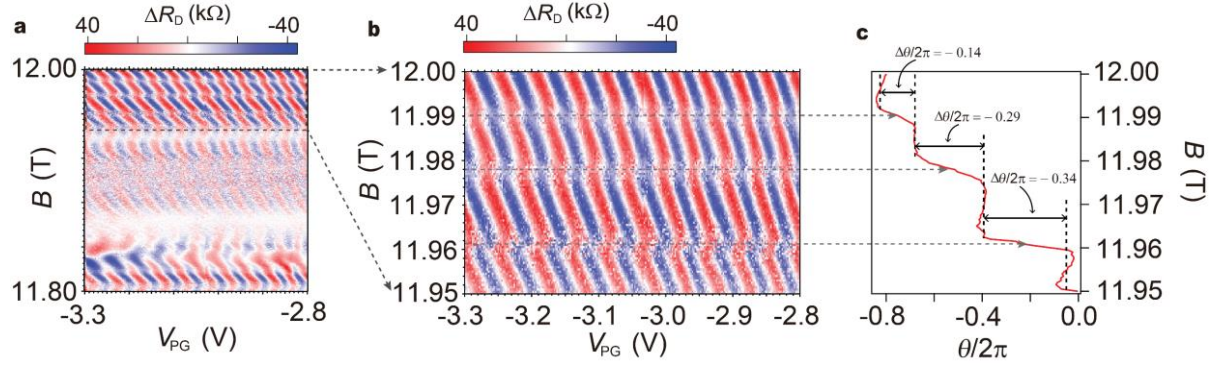

**Fig. S15 | Phase slips. a-b**, Oscillating $\Delta R_D$ in the $B$-$V_{PG}$ plane. (a) Large range of $B$. (b) Zoom-in to small range of $B$ beween 11.95 and 12 T. Three dashed lines indicate where the phase slips occur and point to the corresponding phase slips in Fig. S15c. **c,** scaled phase difference $\theta/2\pi$ extracted from 1D FFT of Fig. S15b as a function of $B$. The AB contribution is subtracted to make plateaus depicted by the dashed lines. The values of the phase slips, $\Delta\theta/2\pi$, are presented between two adjacent plateaus.

**Supplementary Section 9: Extracting the values of phase slips from 1D-FFT**

A pajama plot of $\Delta R_D$ as a function of plunger gate voltage $V_{PG}$ and external magnetic field $B$ is shown in Fig. S16a, the same figure shown in Fig. S15b. To accurately determine the values of the phase slips associated with the quasiparticle inside the interferometer, we employ the Fast Fourier transform method to acquire phase [8]. FFT is performed along the line cut as shown in Fig. S16a with a black dashed line from which $\Delta R_D$ is extracted at a fixed $B$. In this way, we obtain real and imaginary amplitudes associated with that $\Delta R_D$ along that line cut, and the frequency where the amplitude is maximum is determined for further phase calculation. The phase $\theta$ is obtained from inverse tangent function *i.e.*, $\theta(z) = \text{atan2}(Imz, Rez)$, which takes care of the sign of quadrant. To avoid discontinuities while crossing the range from $-\pi$ to $+\pi$, we patch up the phase by a shift of $2\pi$ whenever there is a crossover in this range as shown in Fig. S16b, yielding a smooth curve shown in Fig. S16c. Phase increases linearly until the phase slip emerges as shown in Fig. S16c, which corresponds to the Aharonov-Bohm (AB) phase contribution. To distinguish between AB and quasi-particle phase contributions, constant slopes of all dashed lines representing the AB contribution are averaged out and subtracted from the phase. In this way, we achieve a smooth curve of the phase where the phase slips from anyonic phase contribution become more predominant as shown in Fig. S16d. The same 1D-FFT analysis is performed in Fig. 4e of the main text.

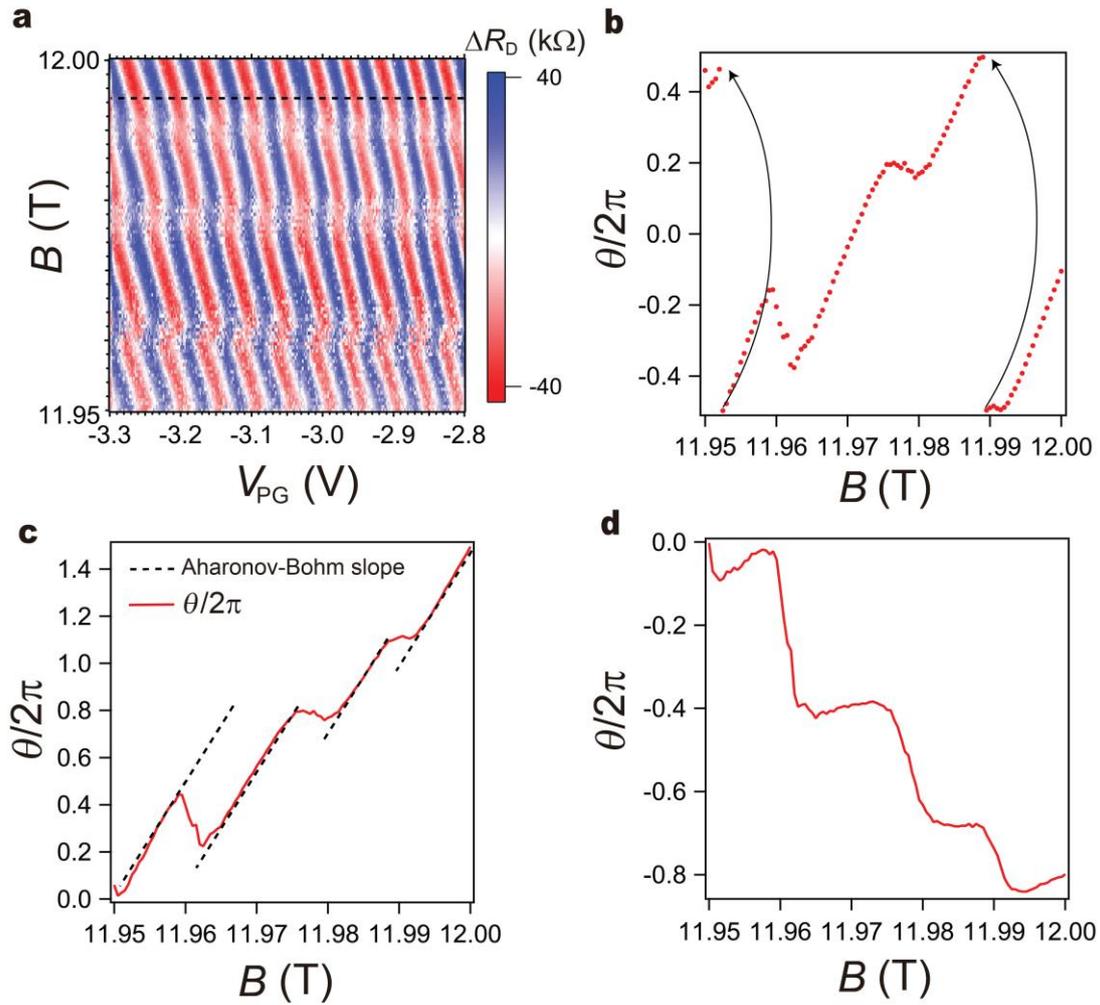

**Fig. S16 | 1D-FFT analysis. a**, Oscillating $\Delta R_D$ as a function of $B$ and $V_{PG}$ for $\nu = 1/3$ at constant density. **b**, The phases accumulated from Fast Fourier Transform for each $B$ value, plotted with the scaled phase $\theta/2\pi$ as a function of $B$. Discontinuities in the phase is avoided by patching up the phase by a shift of $2\pi$ whenever there is crossover of $-\pi$ to $+\pi$, depicted by the black arrows. **c**, The phase after patching up. The black dashed lines represent the constant Aharonov-Bohm phase slope. **d**, The phase with the Aharonov-Bohm slope subtracted to segregate the contribution from the associated phase slips.

**Supplementary Section 10: AB interference at ν = - 1/3 fractional filling**

One of the distinct tunabilities of graphene-based FPI is the fact that the type of interfering charge, either electron or hole, can be controlled by simple gate operations. Now, we demonstrate the possibility of AB oscillations at different filling factor ν = -1/3 where the type of the charge carriers are holes. Figure S17a shows $R_{xy}$ measured at the right side of the FPI while applying 12 T of a magnetic field at fixed 0 V of $V_{BG}$, showing the clear plateaus for IQHE and FQHE. We perform similar measurement performed in Fig. 4d of the main text. We tune to ν = -1/3 in L/RG, and CG regions, and ν = 0 beneath the split gates L/RSG, while measuring oscillating $R_D$ under $I_{SD}$ = 50 pA with $t_L$ ~ 0.95 and $t_R$ ~ 0.75, at constant density regime. Figure S17b shows that AB oscillations measured in $B$ and $V_{PG}$ space have mirror symmetry with respect to $B$ axis, as expected between electron and hole carriers, demonstrating that graphene-based FPI is a promising platform to study both hole and electron types of the interfering charge, and the interplay between them.

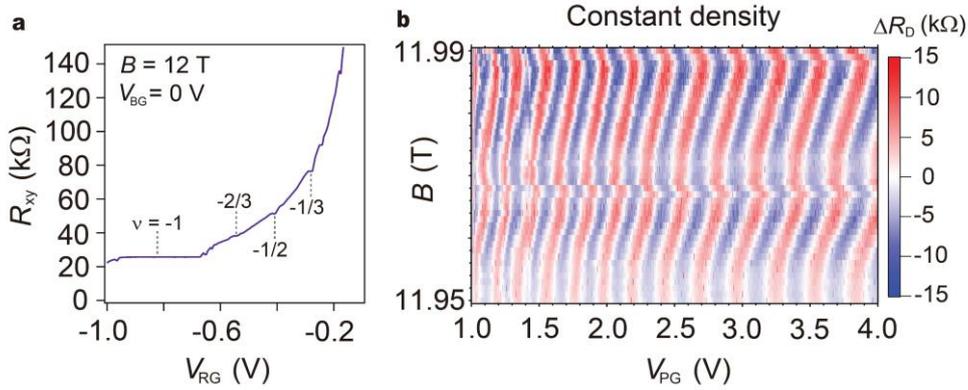

**Fig. S17 | AB interference at a -1/3 fractional filling. a**, $R_{xy}$ measured while applying 2 nA as a function of $V_{RG}$ under 12 T at fixed $V_{BG}$ = 0 V, showing the clear plateau for ν = -1/3. **b**, Oscillating $\Delta R_D$ as a function of $B$ and $V_{PG}$ at constant density condition where the voltages applied on L/RG, and CG are fixed. $V_{BG}$ is set to 0 V.